\documentclass[%
 reprint,
superscriptaddress,
 amsmath,amssymb,
 aps,
]{revtex4-2}

\usepackage{graphicx}% Include figure files
\usepackage{dcolumn}% Align table columns on decimal point
\usepackage{bm}% bold math
\usepackage{multirow}

\begin{document}

\preprint{APS/123-QED}

%\title{Manuscript Title:\\with Forced Linebreak}% Force line breaks with \\
\title{Stars with Plumbing Issues:\\ The Multi-component Ejecta of Common Envelopes}% Force line breaks with \\


\author{Sarah V. Borges}
 \email{villano4@msu.edu}
 \affiliation{%
Department of Physics and Astronomy, Michigan State University, 567 Wilson Road, East Lansing, MI 48824, USA
}%
 \affiliation{%
Department of Physics and Astronomy, University of Wisconsin-Milwaukee, 3135 North Maryland Avenue, Milwaukee, WI 53211, USA
}%
\author{Philip Chang}%
\affiliation{%
Department of Physics and Astronomy, University of Wisconsin-Milwaukee, 3135 North Maryland Avenue, Milwaukee, WI 53211, USA
}%

\date{\today}% It is always \today, today,

\begin{abstract}

Common-envelope evolution (CEE) is one of the major open problems in binary stellar evolution, even though it is the primary channel for forming close binaries with at least one compact object. While CEE is often assumed to lack direct observational constraints, recent connections between CEE and the transient class, luminous red novae (LRNe), as well as the post-AGB systems known as Water Fountains (WFs), provide new opportunities to understand how this event unfolds. Collectively, these observations suggest that envelope ejection during CEE is neither spherically symmetric nor homologous, but rather produces complex ejecta, providing constraints for simulations of CEE. In this work, we present a series of three-dimensional hydrodynamical (non-MHD) simulations of CEE for a 2~$\text{M}_{\odot}$ red-giant-branch (RGB) star and show that the event naturally produces multi-component ejecta including an extended radially expanding shell, a thick circumbinary disk (CBD), and collimated bipolar outflows. Consistent with previous works in the literature, the bipolar outflows in our simulations arise naturally through geometrical constraints imposed by the disk and do not require magnetic fields. While we find that the formation of the collimated outflows can be sensitive to the choice of Riemann solver, they appear across all resolutions and mass ratios tested in this work. We show that the multi-component ejecta formed during CEE simulations naturally reproduce some of the key properties of CEE observations, including the classical bipolar shape seen in some LRNe, WFs, and planetary nebulae.
\end{abstract}

\maketitle

%\tableofcontents

\section{\label{sec:intro} Introduction}
Common-envelope evolution (CEE) is a brief phase in binary stellar evolution during which a giant star engulfs the companion within its envelope. The resulting drag forces extract orbital energy and angular momentum from the system, causing the orbital separation between the companion and the core to shrink. This process is responsible for forming close binaries and increases the probability of future interactions. As a result, CEE plays a key role in the formation rates of various astrophysical systems and events, including X-ray binaries, cataclysmic variables, compact-object mergers, and Type Ia supernovae \citep[for a review see, e.g.,][]{2020cee..book.....I, 2023LRCA....9....2R}.

Despite its significance, CEE remains poorly understood. It is expected to proceed through three main phases: the loss of orbital stability, a dynamical plunge, and a slower spiral-in. Each phase happens on very different timescales: the rapid plunge typically lasts less than a decade, whereas the slower spiral-in can last centuries to millennia. As a result, the dominant energy sources driving the interaction vary between phases. While accretion happens throughout the event, the release of orbital energy due to drag is the main driver of envelope unbinding during the plunge, whereas accretion may contribute significantly to mass loss and orbital decay during the spiral-in.

The historical lack of confirmed observations of ongoing CEE further complicates its modeling. Two main outcomes are possible: (i) the binary merges during the interaction, producing a luminous red nova (LRN), or (ii) the envelope is successfully ejected, leaving behind a close binary system.  Historically, our observational constraints of CEE rely only on post-CEE observations, such as close binary systems that probe the final separation, and bipolar planetary nebulae (PNe), remnants of the ejected envelope. More recently, a group of post-AGBs known as Water Fountains~(WFs) has also been proposed as CEE observations~\citep{2022NatAs...6..275K}.

Among the three phases of CEE, the plunge is currently the best constrained, both observationally through LRNe and numerically through 3D hydrodynamical simulations. Despite this, the amount of envelope ejected and its morphology at the end of this phase remain a matter of debate. In some models, the envelope remains partially bound, and the interaction continues through dynamical friction within an expanded envelope, potentially entering a self-regulated phase in which the released orbital energy is partially radiated away or contributes to the unbinding of the envelope \citep[e.g.,][]{1979A&A....78..167M,2016MNRAS.462..362I}. The envelope is often treated in 1D (approximately spherical), which is difficult to reconcile with the bipolar morphologies observed in post-CEE PNe, LRNe, and WFs. Other studies argue that the envelope is fully ejected during the plunge and that little additional orbital decay happens afterward, either because sufficient orbital energy has already been deposited or because recombination energy contributes significantly to envelope ejection \citep[e.g.,][]{2023MNRAS.526.5365V,2019MNRAS.486.5809P,2020A&A...644A..60S}. However, observations of some post-CEE PNe suggest that mass ejection may continue for thousands of years, rather than happening over a much shorter timescale~\citep[e.g., the PNe NGC~6337,][]{2009ApJ...699.1633G}.
A different possibility is that the plunge naturally leaves behind a thick circumbinary disk (CBD) that governs the subsequent spiral-in phase. The formation of such structures has been reported in several numerical studies \citep[e.g.,][]{2018MNRAS.480.1898C,2019MNRAS.484..631R,2024A&A...691A.244V,2025A&A...697A..68G}. In this scenario, the CBD offers a natural explanation for two open problems. First, it allows for continued orbital evolution after the plunge. Although recombination energy may be sufficient to unbind the envelope, plunge simulations often predict post-CEE binary separations larger than those inferred from observed post-CEE binaries \citep[see discussion in][]{2026ApJ...998...34B}, suggesting that interactions after the end of the plunge phase remain important. Second, it can allow for the formation and longevity of bipolar outflows.

Several mechanisms have been proposed to explain the bipolar morphologies of PNe. Rotation and magnetic fields in single stars are unlikely to be sufficient, since most of the angular momentum is lost during the AGB phase~\citep{2014ApJ...783...74G,2016ApJ...823..142G}, which has led to the suggestion that bipolar PNe are products of binary interactions, including CEE~\citep[e.g.,][]{2017NatAs...1E.117J}. Models of post-CE nebulae have attributed the bipolar shape to (i) a fast wind from the remnant core interacting with the flattened CE ejecta~\citep{2018ApJ...860...19G,2020MNRAS.497.2855Z}; (ii) magnetically driven winds and jets~\citep{2020ApJ...893..150G,2021ApJ...914..111G,2022A&A...660L...8O,2025A&A...698A.133V}; and (iii) purely hydrodynamical outflows escaping through a centrifugally evacuated polar funnel~\citep{2025A&A...697A..68G}. These models rely on one of two ingredients: magnetic fields, or a dense equatorial structure that confines the outflow like a de Laval nozzle~\citep{1981ApJ...247..152I,2025ApJ...980..119B}. 

Observations suggest that both may be needed: W43A and HD~101584 show both a wide-angle collimated outflow and a narrower jet~\citep{2020ApJ...890L..14T,2019A&A...623A.153O}, and IRAS~16342$-$3814 has an outer bipolar structure carved by a precessing jet~\citep{2005ApJ...622L..53S}. A similar picture emerges in some evolved PNe, which combine extended bipolar lobes with signatures of point-symmetric jets~\citep[e.g., NGC~2818 and the Red Spider Nebula,][]{2024MNRAS.530.3327D,2025ApJ...993...79K}. These morphologies may reflect the interplay between two components: a wide-angle outflow collimated by an equatorial structure, and a narrower, precessing jet, which can significantly reshape the bipolar structure~\citep{2000ApJ...544..336G}. Since jets are generally associated with accretion, a CBD likely plays a double role: collimating the wide-angle outflow and providing a reservoir of material to sustain accretion and, consequently, jets.

Several questions remain open. First, it is unclear whether large-scale wide-angle collimated outflows form self-consistently during CEE and shape the ejecta on large scales. \citet{2025A&A...697A..68G} showed that such outflows can arise from pure hydrodynamics but \citet{2025RAA....25b5023S} argued that they are too weak and would be deflected before forming the bipolar lobes of PNe. Second, it is unclear why some simulations find a CBD while others do not~\citep[e.g.,][]{2016MNRAS.462..362I,2023MNRAS.526.5365V}, and whether this difference arises from numerics or from different initial parameters. Third, the identification of a CBD is often based on the presence of equatorial, rotating gas around the binary. Only a few studies have tested whether the equatorial gas is actually a stable, rotationally supported disk~\citep[e.g., for a red supergiant donor,][]{2024A&A...691A.244V}. Finally, the CBD is not the only structure left behind by the plunge, and the other components of the ejecta, such as an outer shell expanding adiabatically like that of the Boomerang Nebula~\citep{2017ApJ...841..110S}, have received less attention.

We therefore revisit this problem with three goals: (i) to describe the multi-component morphology of the envelope at the end of plunge, and compare it with observations of LRNe, WFs, and PNe; (ii) to understand how the mass ratio, corotation and numerics affect the formation of a CBD; and (iii) to provide initial conditions for future simulations of the spiral-in. Ideally, a single simulation would model both the plunge and the subsequent spiral-in. However, the spiral-in will likely require additional physical processes, such as cooling and accretion, and maintaining high spatial resolution near the binary for particle-based simulations becomes increasingly challenging as the envelope expands and mass is progressively unbound. It is therefore advantageous to simulate these phases separately, with a careful description of the post-plunge envelope serving as the starting point for the spiral-in.

For that, we performed ten three-dimensional hydrodynamical simulations of CEE with a red-giant-branch (RGB) donor, continuing the work of \citet{2026ApJ...998...34B}, where we discussed the final separation at the end of the plunge. We first present our simulations, how numerics affect the formation of a CBD, and how the system evolves from the plunge to the spiral-in phase~(Secs.~\ref{sec:methods} and~\ref{sec:spiral_in}). We then describe the resulting multi-component ejecta, composed of an outer shell, a torus, a CBD, and collimated bipolar outflows~(Sec.~\ref{sec:components}), and compare their morphology and kinematics with those observed in LRNe, WFs, and bipolar PNe~(Sec.~\ref{sec:observations}). Finally, we discuss how observations can guide the study of the poorly constrained spiral-in phase of CEE~(Sec.~\ref{sec:discussion}).

\section{Simulating a CEE}
\label{sec:methods}

In this work, we analyze a subset of the 3D CEE simulations presented in \citet{2026ApJ...998...34B}. In addition, we include a new simulation with a lower temperature floor and three simulations with different resolutions. Table~\ref{tab:our_simulations} summarizes the parameters of the simulations analyzed here. The simulations were performed with the moving-mesh hydrodynamics code \texttt{MANGA}~\citep{2017MNRAS.471.3577C}, which is a module of the SPH code \texttt{ChaNGa}~\citep{jetley2008,jetley2010,2015ComAC...2....1M}. Other than CEE, \texttt{MANGA}~has previously been used to simulate stellar mergers~\citep{2017MNRAS.471.3577C}, tidal disruption events~\citep{2021MNRAS.501.1748S}, TZO formation~\citep{2025ApJ...993...61W} and classical novae~\citep{2026PhRvD.113l3023N}. 

For a more complete discussion of the methods, please refer to \citet{2026ApJ...998...34B}. In short, the stellar model was generated with \texttt{MESA}~\citep{2011ApJS..192....3P,2013ApJS..208....4P,2015ApJS..220...15P,2018ApJS..234...34P,2019ApJS..243...10P,2023ApJS..265...15J}, evolved from the pre-main sequence to the red-giant branch (RGB). We considered a $2~\mathrm{M}_\odot$ RGB primary with a radius $R_1$ of $53~\mathrm{R}_\odot$. The core of the giant and the companion (initially placed at the stellar surface) were replaced by softened point masses, and the envelope considers an adiabatic equation of state ($\gamma = 5/3$). The softening radius of the companion is given by $(M_2/\mathrm{M}_\odot)\mathrm{R}_\odot$, where $M_2$ is the secondary mass, and that of the core is 0.5~$\mathrm{R}_\odot$. The number of particles is 2 million, with 1.2 million for the star. The simulations were performed considering the \texttt{HLLC} solver, a floor temperature of $10^4$~K, and corotation, but we also have six additional simulations to check such choices: one with the \texttt{HLL} solver, three with lower and higher resolutions (0.6, 1.6, and 2.1 million star particles), one with a floor temperature of $10^2$~K, and a final one without corotation~(all for $q=0.5$). Most cases ran for $1\,500$ days, except for Q050-NCR~($1\,000$~days), Q050-T100~($5\,000$~days), Q050-0.6M~(500~days), Q050-1.6M~(500~days), and Q050-2.1M~(500~days). The impact of numerics (Riemann solver and resolution) is discussed in Appendix~\ref{sec:solver}. In short, the solver does impact the formation of a CBD, which shows up only in the \texttt{HLLC} runs, but the resolution does not. The change in resolution impacts the final separation and how fast the disk is formed.

\begin{table}[htbp]
\centering
\caption{Parameters of our CEE simulations. All runs use a $2~\text{M}_\odot$ RGB primary with core mass $M_{1,\text{c}} = 0.37~\text{M}_\odot$ and radius $R_1 = 53~\text{R}_\odot$. The mass ratio is $q = M_2/M_{1,\text{t}}$, where $M_{1,\text{t}}$ is the total~(core+envelope) mass of the primary, $a_{\rm f}$ is the semi-major axis at the end of the plunge as defined in \citet{2026ApJ...998...34B}, $N_\star$ is the number of star particles, $T_{\rm floor}$ is the floor temperature, and Cor.\ indicates whether the envelope initially corotates with the binary.}
\begin{tabular}{ccccccc}
\hline
Run & $q$ & $a_{\rm f}$~($\text{R}_\odot$) & $N_\star$~($10^6$) & Solver & $T_{\rm floor}$~(K) & Cor. \\
\hline
Q025      & 0.25 & 3.5  & 1.2 & \texttt{HLLC} & $10^4$ & Yes \\
Q050      & 0.50 & 6.9  & 1.2 & \texttt{HLLC} & $10^4$ & Yes \\
Q075      & 0.75 & 11.2 & 1.2 & \texttt{HLLC} & $10^4$ & Yes \\
Q100      & 1.00 & 16.7 & 1.2 & \texttt{HLLC} & $10^4$ & Yes \\
\hline
Q050-HLL  & 0.50 & 3.8  & 1.2 & \texttt{HLL}  & $10^4$ & Yes \\
Q050-T100 & 0.50 & 7.0  & 1.2 & \texttt{HLLC} & $10^2$ & Yes \\
Q050-NCR  & 0.50 & 5.7  & 1.2 & \texttt{HLLC} & $10^4$ & No  \\
Q050-0.6M & 0.50 & 8.4  & 0.6 & \texttt{HLLC} & $10^4$ & Yes \\
Q050-1.6M & 0.50 & 7.1  & 1.6 & \texttt{HLLC} & $10^4$ & Yes \\
Q050-2.1M & 0.50 & 6.1  & 2.1 & \texttt{HLLC} & $10^4$ & Yes \\
\hline
\end{tabular}
\label{tab:our_simulations}
\end{table}

\begin{figure*}[htbp]
\parbox{\textwidth}{%
\includegraphics[width=0.95\textwidth]{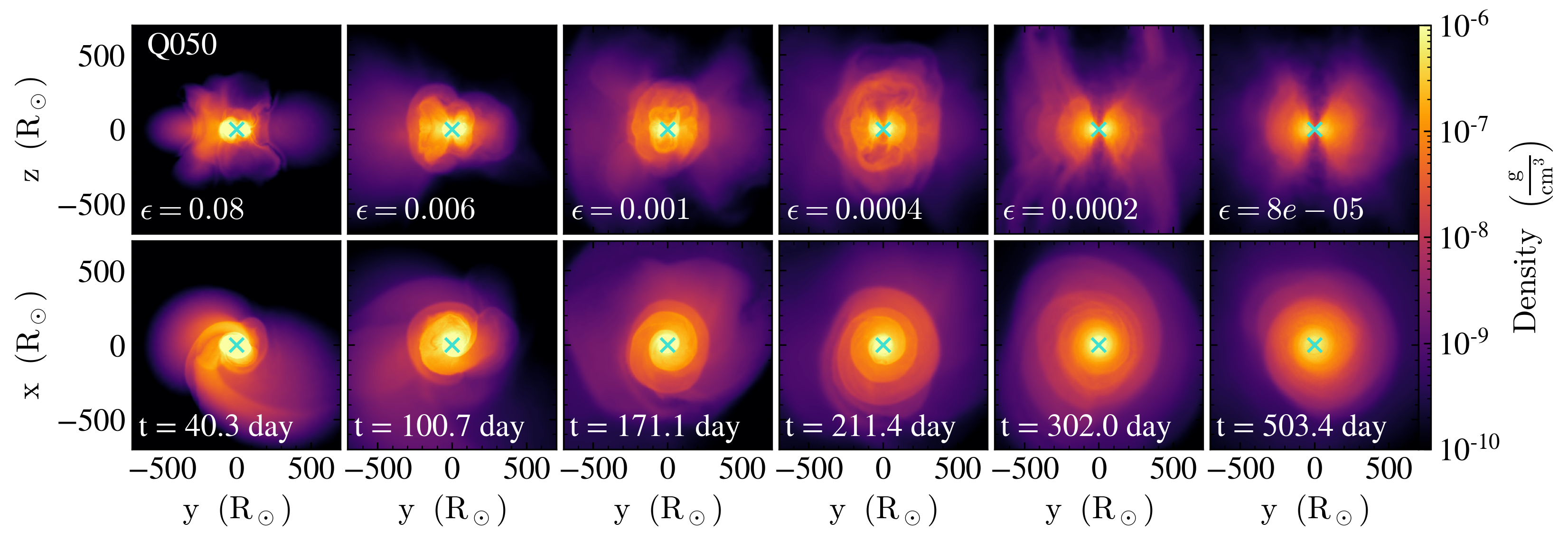}\\
}

\caption{Density distribution for a $2~\text{M}_\odot$ RGB star in a binary system with mass ratio $q=0.5$ for the Q050 run, \texttt{HLLC} Riemann solver, 1.2 million particles. Each panel shows both $y$-$x$ and $y$-$z$ views at times $t = 40.3$, $100.7$, $171.1$, $211.4$, $302$, and $503.4$ days. Each panel has the $\epsilon$ at that given time. The cyan '$\times$' marks the center of mass of the core+companion duo.} 
\label{fig:projection_plots}
\end{figure*}

\begin{figure}[htbp]
\includegraphics[width=0.45\textwidth]{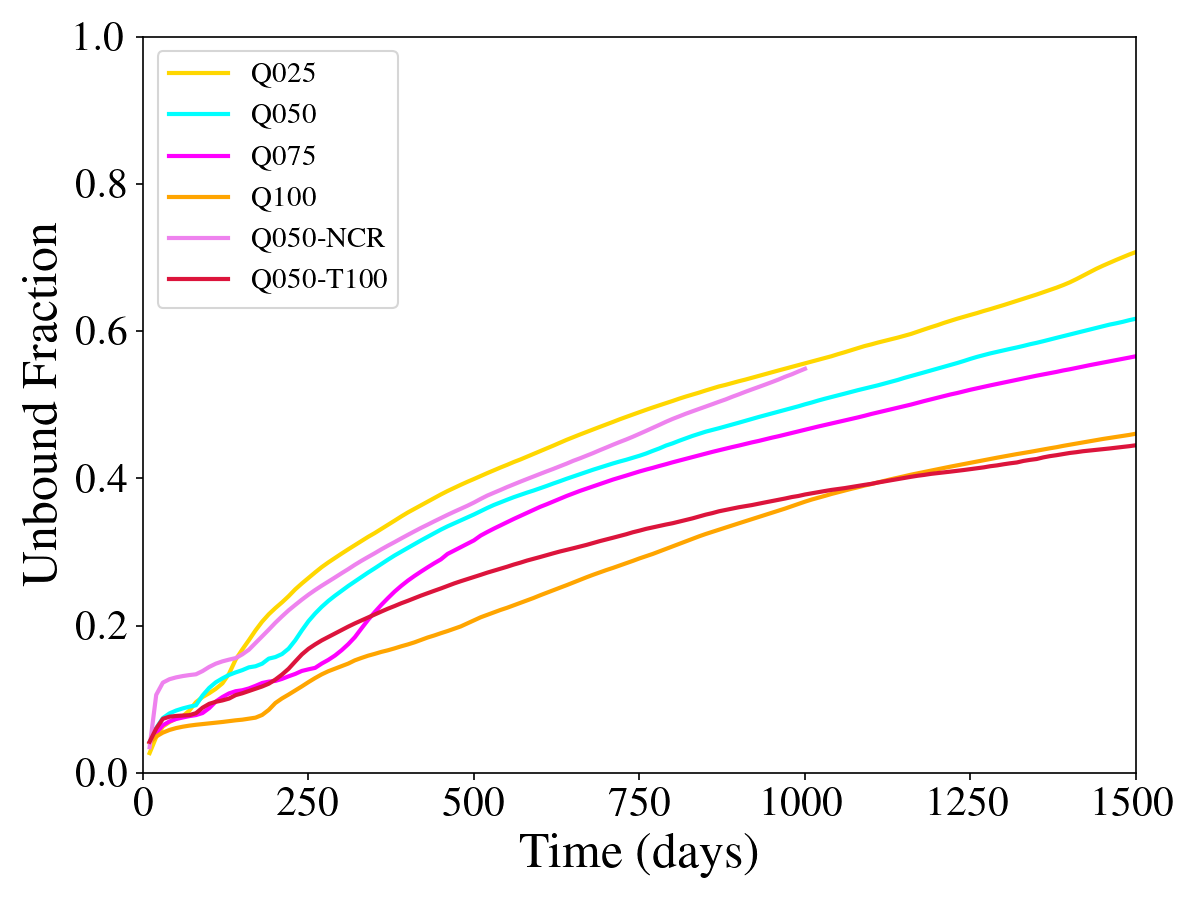}

\caption{ Total unbound fraction over time~(Appendix~\ref{sec:solver} discusses how the unbound fraction is estimated). The unbound fraction for the run  Q050-HLL is shown in figure~\ref{fig:unbound_sep}. The colors are yellow, cyan, magenta, orange, pink and red, for the runs Q025, Q050, Q075, Q100, Q050-NCR, and Q050-T100, respectively. Those colors will be used for the same runs throughout the paper.} 
\label{fig:unbound_all}
\end{figure}

\section{The onset of the spiral-in phase}
\label{sec:spiral_in}

Describing the envelope at the end of the plunge first requires defining when the plunge ends, since this determines when the multi-component structure of the ejecta becomes a valid description of the system. In general, the transition between the plunge and spiral-in is identified using the condition:

\begin{equation} \label{eq:threshold}
|\dot{a} P / a| < \epsilon,
\end{equation}

\noindent where $a$ is the semi-major axis, $P$ is the orbital period between the core and the companion, and $\epsilon$ is a threshold value, typically 0.1 or 0.01. This criterion was introduced as a practical way to distinguish the rapid orbital decay of the plunge phase from the subsequent slower spiral-in. However, the physical meaning of this boundary remains debated \citep[see, e.g.,][]{2020cee..book.....I}. If the spiral-in phase is interpreted as a self-regulated continuation of drag-driven evolution, the distinction may be largely operational, reflecting a gradual decrease in drag due to the envelope puffing-up rather than a sharp boundary in the underlying physical regime. In contrast, if a CBD forms, the transition between phases is not defined only by a smaller rate of orbital decay but by a qualitative change in the geometry and relevant physics of the system. From this perspective, morphological indicators can provide a more physically motivated definition of the change in regime. In our \texttt{HLLC} runs, we define the beginning of spiral-in as the moment in which a CBD is fully formed, the central binary resides in a low density cavity, and the polar outflows emerge.

\begin{figure*}[htbp]
\includegraphics[width=0.47\textwidth]{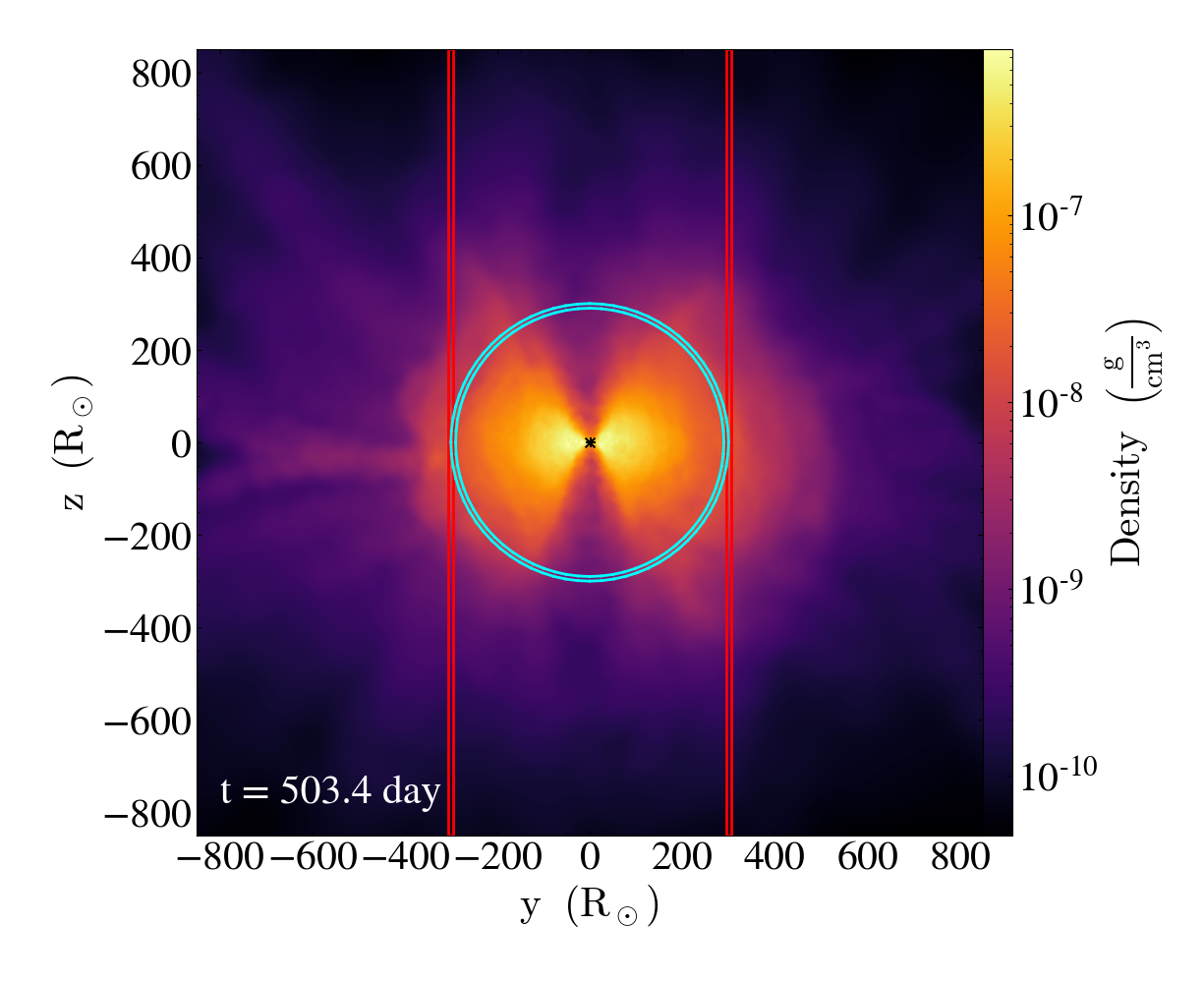}
\includegraphics[width=0.47\textwidth]{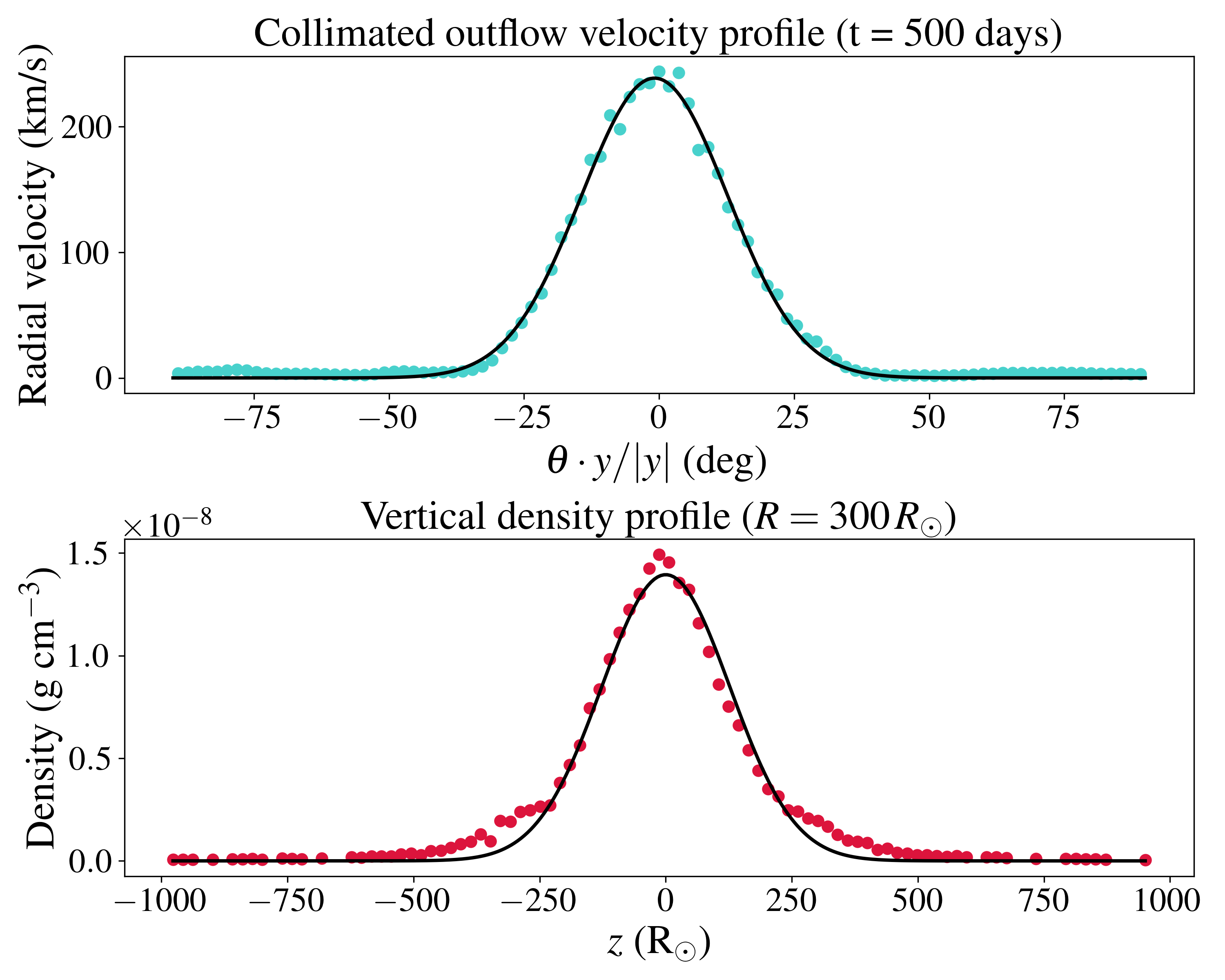}

\caption{{\it Left:} Density distribution for a $2\,\text{M}_\odot$ RGB star in a binary system with mass ratio $q=0.5$~(Q050-T100) in the $y$--$z$ plane at $t=503.4$ days. The cyan circle marks the spherical slice with radius $300\,R_\odot$ around the center in which the radial-velocity Gaussian is calculated over $\theta$ (see Sec.~\ref{sec:outflows}). The red lines mark the slice of the cylinder with radius $300\,R_\odot$ in which the density Gaussian is calculated over $z$~(see Sec.~\ref{sec:density}). 
{\it Right:} Gaussian fits for the velocity distribution over $\theta$ (top) and the density distribution over $z$ (bottom).}
\label{fig:gaussian}
\end{figure*}

For run Q050, for instance, this happens at approximately 200 days, corresponding to $\epsilon \sim 10^{-3}$~(see Fig.~\ref{fig:projection_plots}). This suggests that the commonly adopted threshold ($\epsilon = 0.01$) identifies a boundary that precedes the transition associated with disk formation. Although we still find that defining the separation between phases through equation~\ref{eq:threshold} is advantageous as a standardized way to compare simulations, we advocate adopting a stricter value of $\epsilon = 10^{-3}$ (or even $10^{-4}$) to allow the disk and outflows time to form. Choosing this stricter definition is also beneficial for multi-phase modeling of CEE, because it allows the event to transition to its new disk-driven stage. A similar discussion was raised in our previous work~\citep{2026ApJ...998...34B}, where we were interested in the final separation at the end of plunge and found that considering the end of the plunge at $\epsilon=10^{-3}$ was a better transition~(we briefly discuss this orbital separation argument for a smaller $\epsilon$ in Fig.~\ref{fig:unbound_sep}). In this paper, we show that this later transition also agrees with the geometry of the envelope.

A stricter threshold is also consistent with the idea that the plunge phase proceeds faster than the spiral-in phase. It is therefore better for the threshold to extend slightly into the spiral-in phase rather than into the plunge. For example, if the physical transition happens at 200 days, selecting 250 days as the beginning of the spiral-in phase is better than selecting 150 days, since the state of the system at 250 days is more similar to that at 200 days than the state at 150 days.

\begin{figure*}
    \centering
    \includegraphics[width=0.97\linewidth]{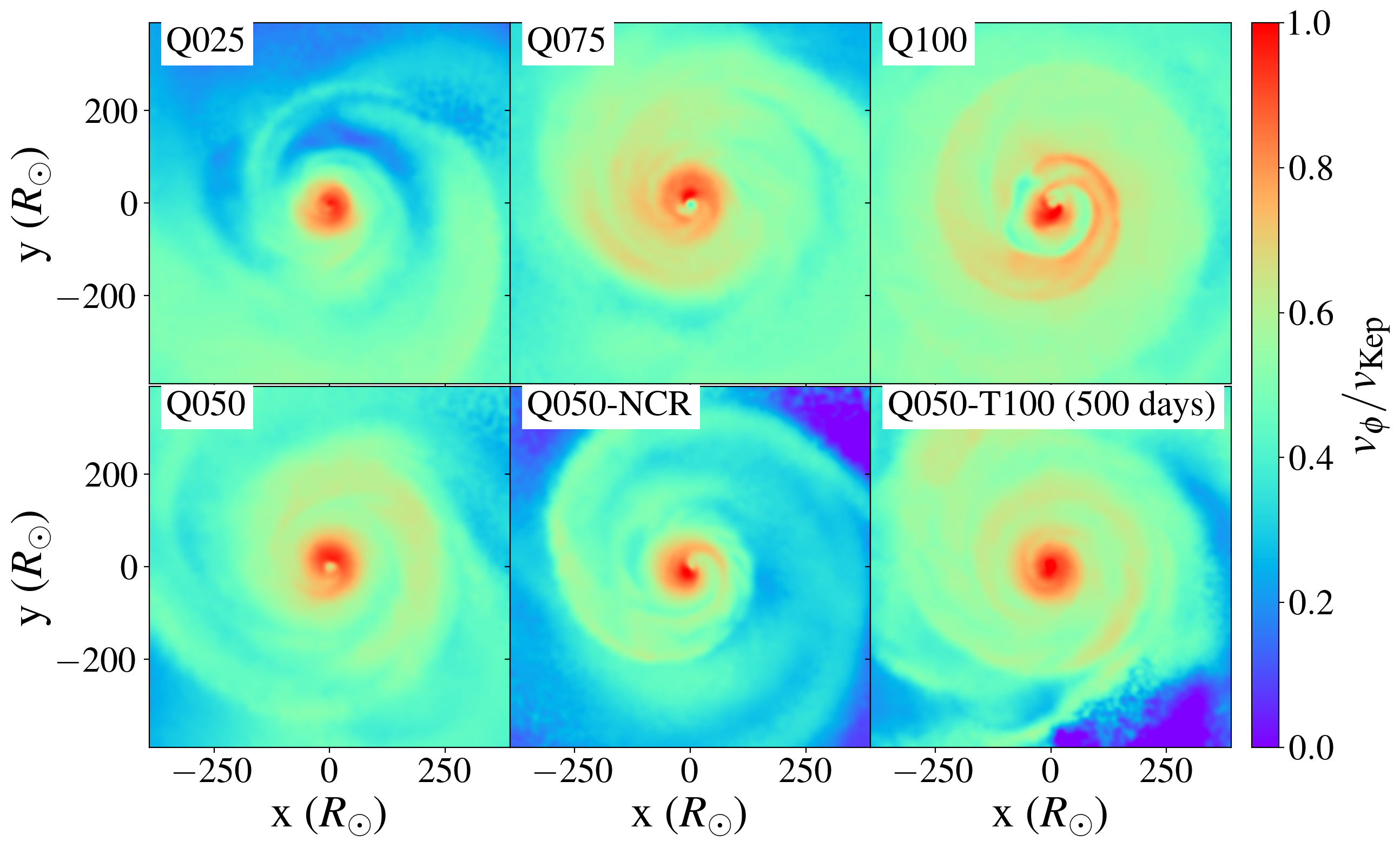}
\caption{Ratio $v_{\phi}/v_{\rm kep}$ in the $x$--$y$ plane at $t = 500$ days for the runs Q025, Q075, Q100, Q050, Q050-NCR, and Q050-T100, where $v_{\rm kep} = [G(M_{1,\rm c} + M_2+ M_{\rm d,enc})/R]^{1/2}$ where $M_{\rm d,enc}$ is the mass of the disk enclosed in R, estimated from Eq.~\ref{eq:mass_disk} for $R_{\rm out} = R$.}
    \label{fig:keplerian}
\end{figure*}

\begin{figure*}
\includegraphics[width=0.47\textwidth]{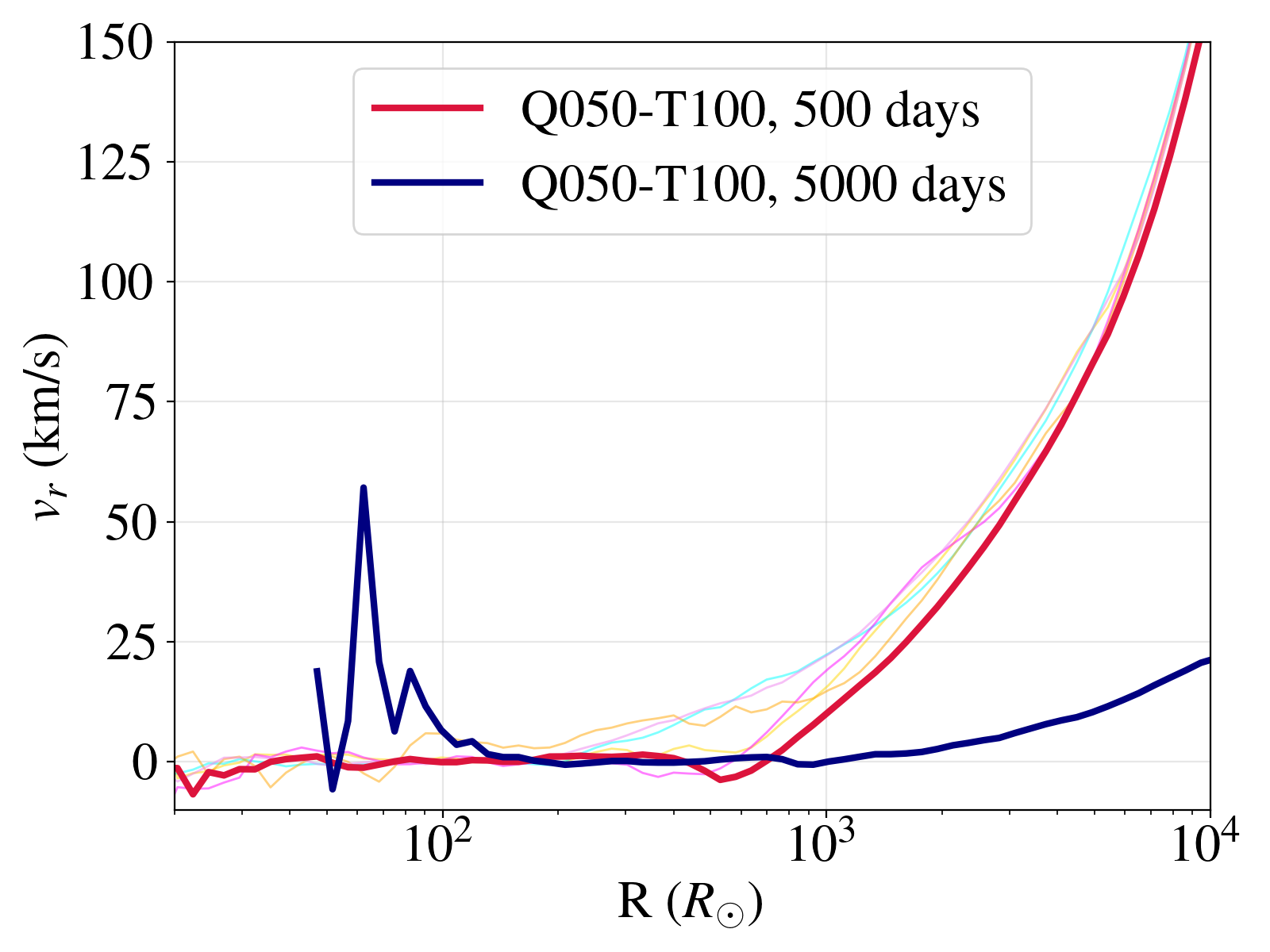}
\includegraphics[width=0.47\textwidth]{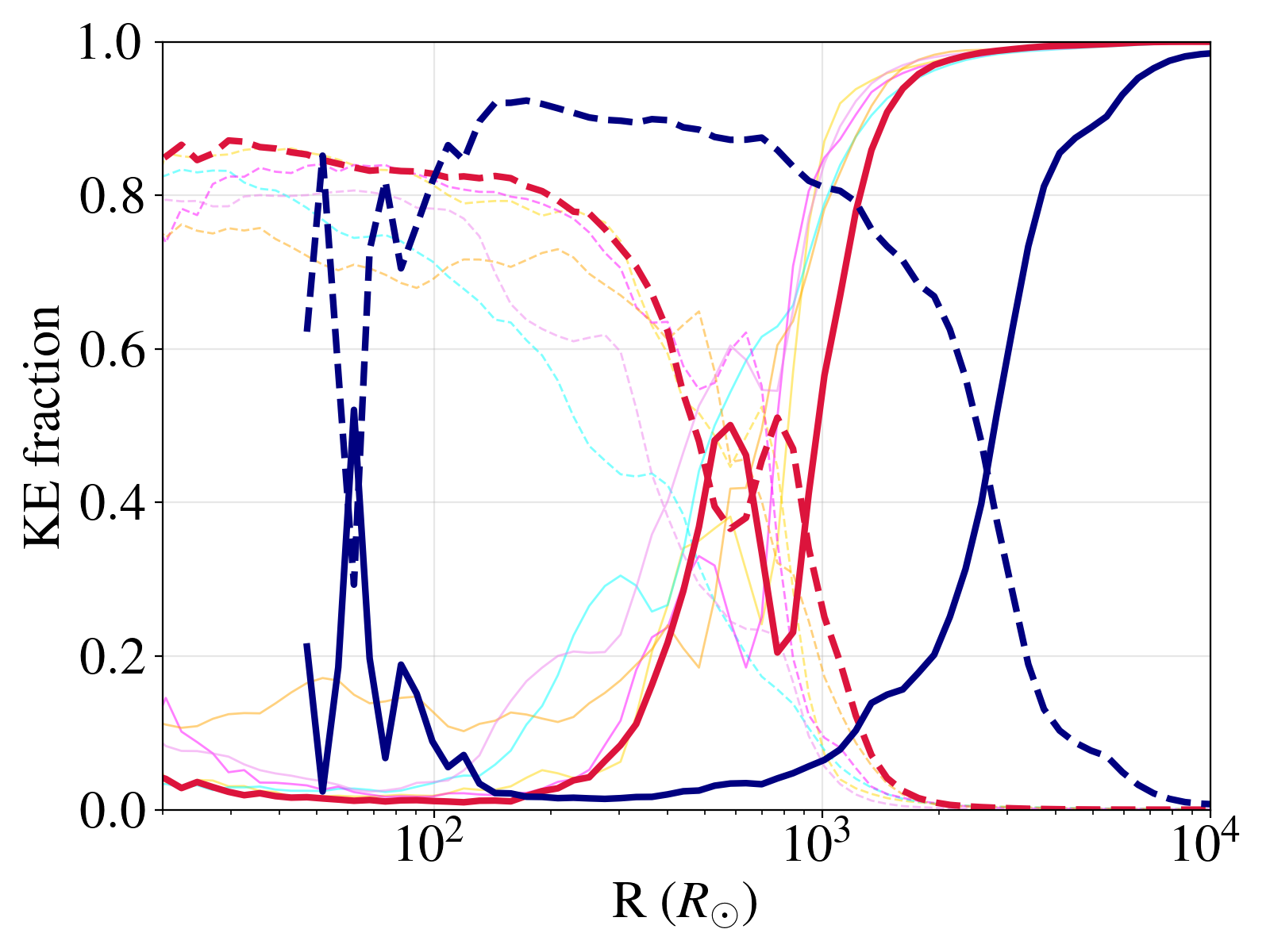}
\caption{{\it Left:} Radial velocity ($v_r$) for $30^{\circ}<\theta<150^{\circ}$~(to exclude the region of the outflow) and  as a function of radius for the runs Q025, Q050, Q075, Q100, and Q050-NCR at $t = 500$ days (faint lines, the colors per simulation run are the same as Fig~\ref{fig:unbound_all}), and for Q050-T100 at $t = 500$ and $5\,000$ days (bright red and navy lines, respectively). {\it Right:} Fraction of kinetic energy in the azimuthal (${\phi}$, dashed lines) and radial ($r$, solid lines) components as a function of radius for the same $\theta$ constraint.}

\label{fig:radial_velocity}
\end{figure*}

\begin{figure}
\includegraphics[width=0.98\linewidth]{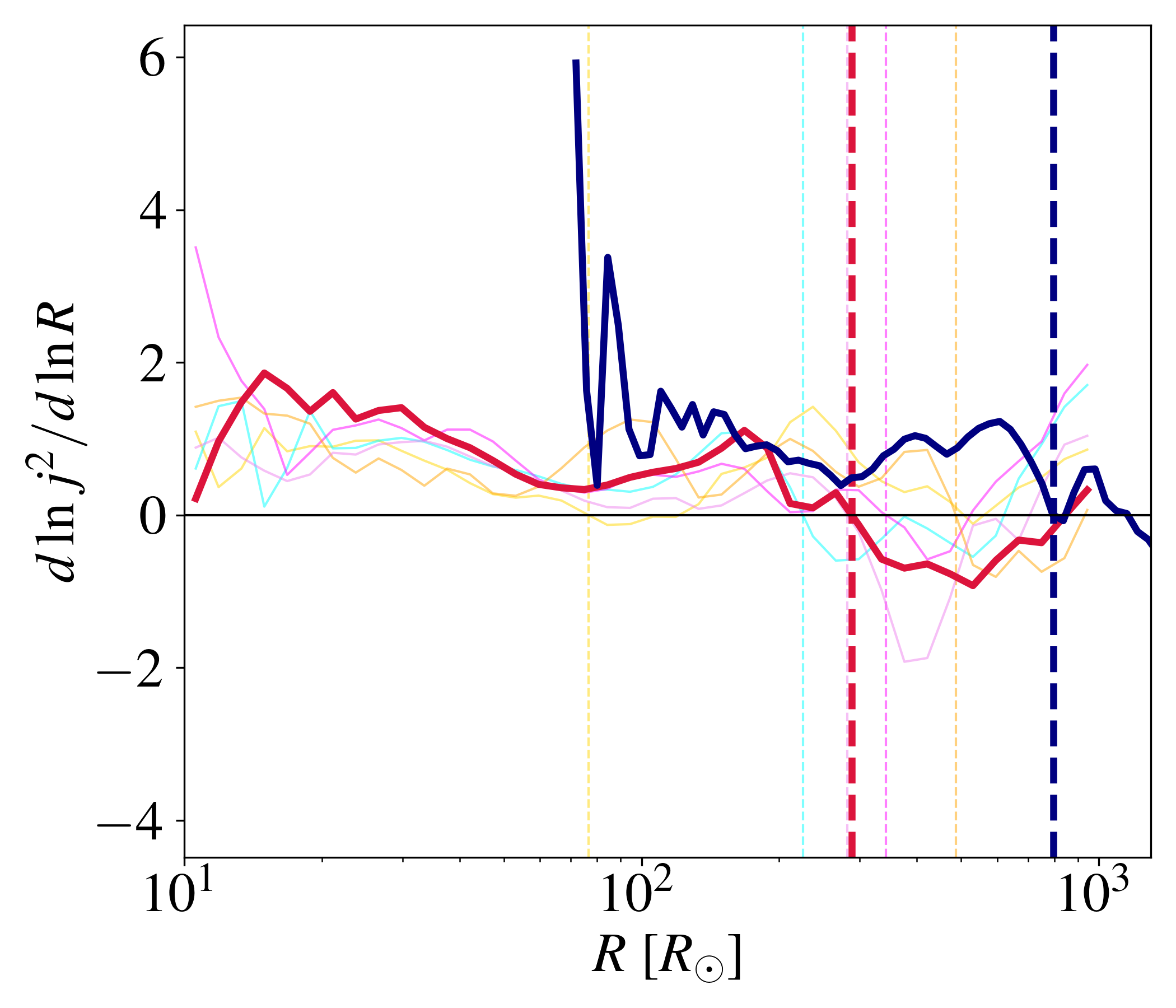}
\caption{Rayleigh stability of the bound equatorial gas ($30^\circ < \theta < 150^\circ$), shown as the logarithmic derivative of the mass-weighted specific angular momentum, $d\ln j^2/d\ln R$, as a function of cylindrical radius, at 500 days (faint lines, with the same colors as Fig.~\ref{fig:unbound_all}) and for Q050-T100 at 500 and 5\,000 days (bright red and navy lines, respectively). The horizontal black line marks marginal stability. Vertical dashed lines, in the color of each run, mark the first radius at which the profile becomes Rayleigh-unstable: $\sim 75\,R_\odot$ (Q025), $\sim 225\,R_\odot$ (Q050), $\sim 285\,R_\odot$ (Q050-NCR), $\sim 290\,R_\odot$ (Q050-T100), $\sim 340\,R_\odot$ (Q075), $\sim 490\,R_\odot$ (Q100), and $\sim 790\,R_\odot$ (Q050-T100 at 5\,000 days).}
\label{fig:angular_momentum}
\end{figure}

\section{Components of the envelope}
\label{sec:components}

\subsection{Thick disk}

\begin{figure*}
    \centering
    \includegraphics[width=0.45\linewidth]{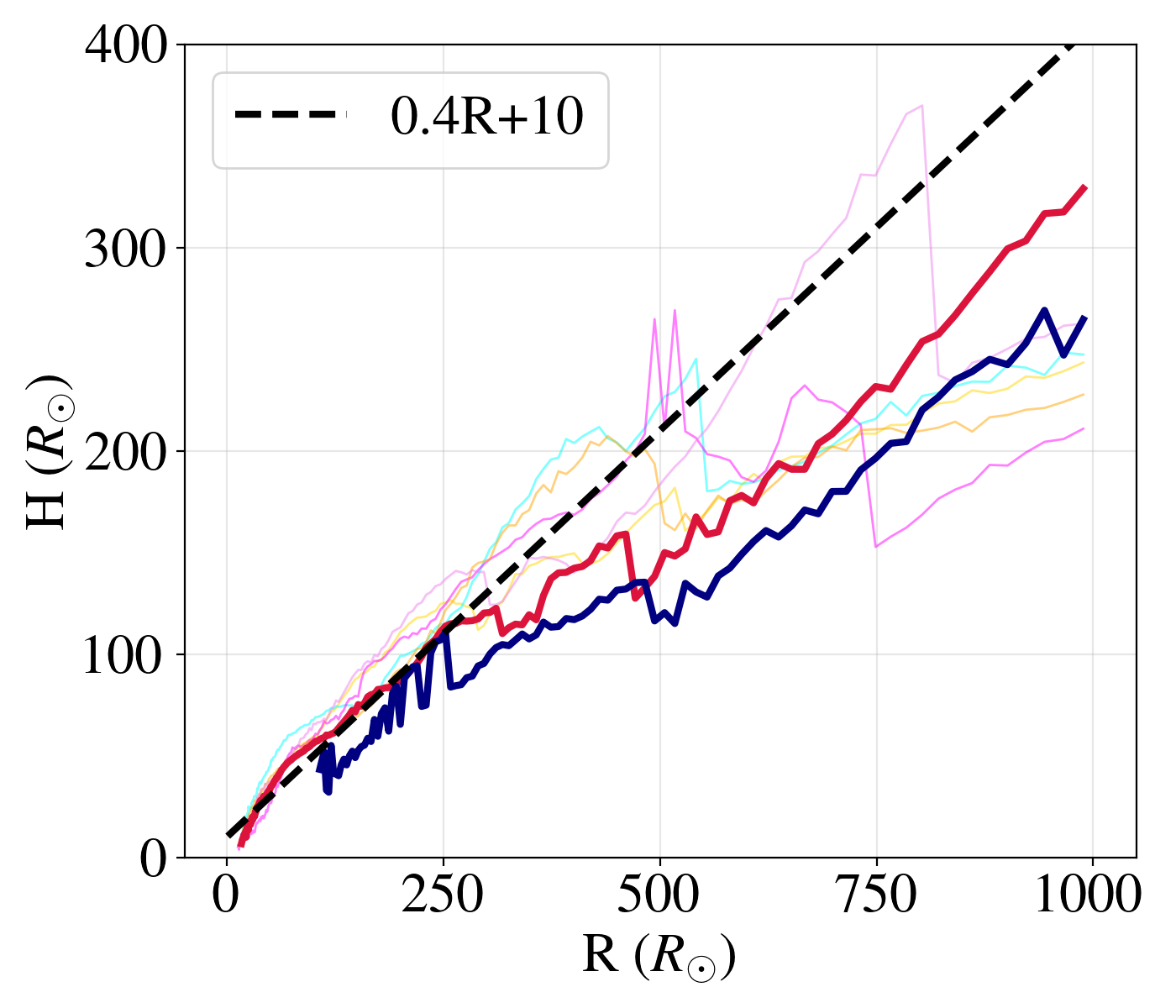}
    \includegraphics[width=0.45\linewidth]{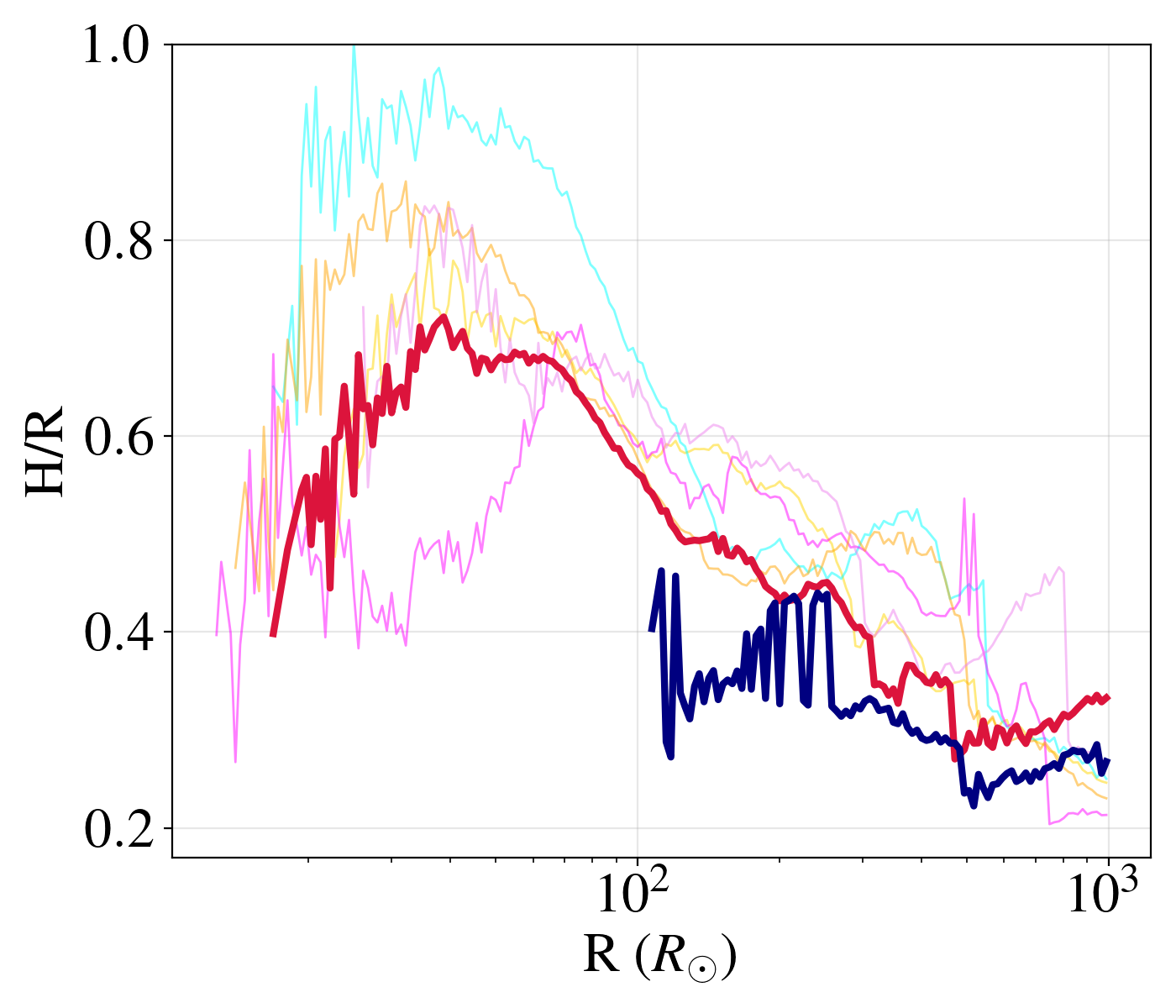}
    \caption{{\it Left:} Disk height $H$ as a function of radius for the runs Q025, Q050, Q075, Q100, and Q050-NCR at $t=500$ days (faint lines, the colors per simulation run are the same as Fig~\ref{fig:unbound_all}), together with Q050-T100 at $t=500$, and $5\,000$ days (red and blue bright lines, respectively). The black dashed line shows the best linear fit to the Q050-T100 profile at $t=500$ days. {\it Right:} Disk thickness $H/R$ as a function of radius for the same runs.}
    \label{fig:height}
\end{figure*}

\begin{figure*}
    \centering
    \includegraphics[width=0.45\linewidth]{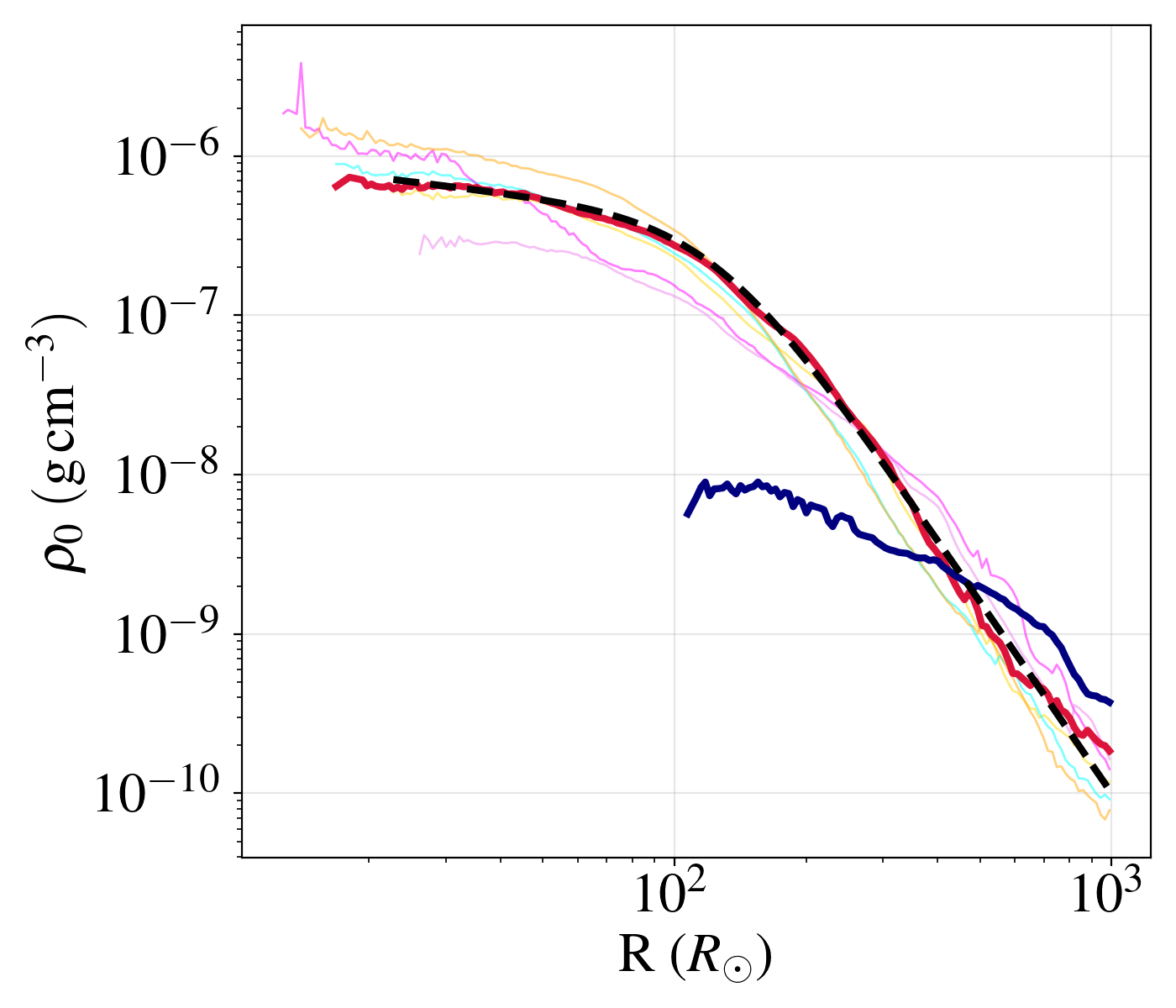}
    \includegraphics[width=0.45\linewidth]{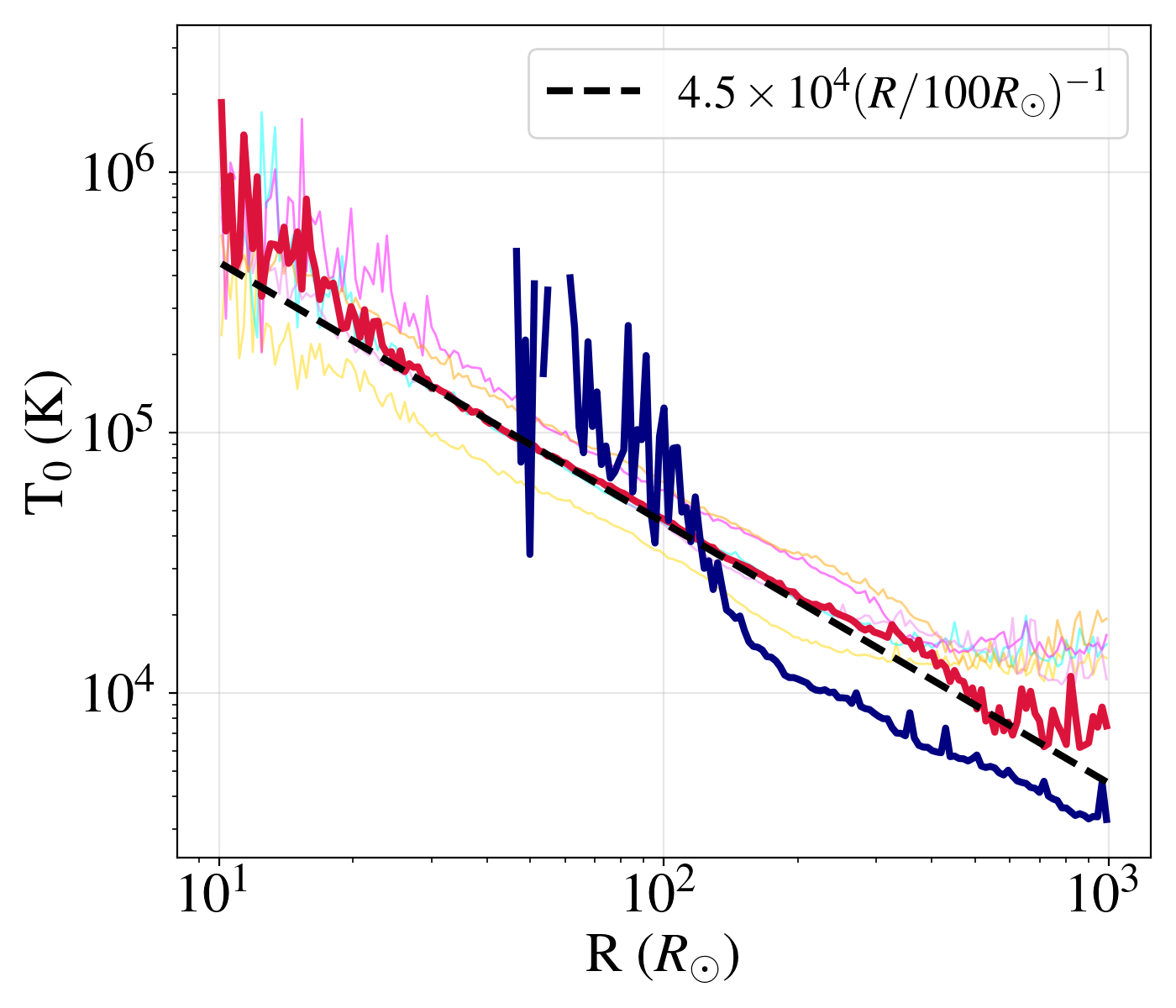}
    \caption{{\it Left:} Midplane density $\rho_0$ as a function of radius for the runs Q025, Q050, Q075, Q100, and Q050-NCR at $t=500$ days (faint lines, the colors per simulation run are the same as Fig~\ref{fig:unbound_all}), together with Q050-T100 at $t=500$, and $t=5\,000$ days (red, and blue bright lines, respectively). The fit from Eq.~\ref{eq:central_density} is shown in a black dashed line. {\it Right:} Midplane temperature $T_0$ as a function of radius for the same runs. In all cases, the temperature approximately follows $T_0 \propto R^{-1}$.}
    \label{fig:density_temperature}
\end{figure*}

We analyzed the runs Q025, Q050, Q075, Q100, and Q050-NCR at $t = 500$~days, and Q050-T100 at $t = 500$ and $5\,000$~days. We have chosen 500 days and not 200 days to allow the formed CBD to adjust, and the quantitative analysis below relies on this snapshot. We extend Q050-T100 to $5\,000$~days primarily to test the survivability of the CBD rather than to predict all the detailed properties of the system at late times. Over this decade-long timescale, physics not included in our simulations, such as cooling and accretion onto the central binary, are expected to become important and may change the structure of the disk, as well as interaction with the central binary.

The analysis focuses on the run Q050-T100, with the other runs shown for comparison, because its lower temperature floor allows the gas to cool below $10^4$~K, so the outer disk and ejecta are not artificially heated by the floor. Moreover, the late-time growth of the unbound fraction depends on the temperature floor, which adds internal energy to the cold, marginally bound outer ejecta and artificially increases the unbound fraction, which can be seen in Fig.~\ref{fig:unbound_all}, in which the run Q050-T100 has the lowest unbound fraction by the end of the $1\,500$~days.

As we will discuss in the next few sections, at 500 days, the midplane temperature of the CBD remains above the default floor of $10^4$\,K out to $R \sim 500\,R_\odot$, beyond the edge of the disk and its Rayleigh-stable region, so the floor does not affect the CBD. To test this, we used the run Q050-T100~(which has a lower floor temperature of 100~K) as a control simulation. At 500 days, the CBDs in runs Q050 and Q050-T100 have similar midplane densities, temperature and scale heights. Because the CBD cools over time, making this artificial heating increasingly important, Q050-T100 is the only run evolved to longer timescales.

\subsubsection{Stability and size of the disk}
\label{sec:disk_size}

Following \citet{2024A&A...691A.244V}, we use the Rayleigh criterion as a diagnostic of rotational stability. In Fig.~\ref{fig:angular_momentum} we show the Rayleigh stability of the bound equatorial gas. The Rayleigh criterion requires $dj^2/dR > 0$, where $j$ is the mass-weighted specific angular momentum. Since $d\ln j^2/d\ln R = (R/j^2)\,dj^2/dR$ has the same sign, we show this dimensionless logarithmic derivative, which allows for a better scale in the plot. All runs are Rayleigh-stable in their inner regions, out to a radius that increases with mass ratio, from $\sim225\,R_\odot$ for Q050 to $\sim490\,R_\odot$ for Q100 at 500 days. For Q025, the profile is stable only in a narrow region, dipping slightly below zero already at $\sim75\,R_\odot$. We adopt the radius at which the Rayleigh criterion first fails as the outer edge of the rotationally supported CBD. Comparing with Fig.~\ref{fig:radial_velocity}, the CBD does not connect directly to the radially expanding component: between them lies a torus of bound, mostly equatorial gas that is not rotationally supported (Sec.~\ref{sec:shell}). For Q050-T100 at 500 days, the disk ends at $\sim290\,R_\odot$, while the radial kinetic energy dominates only beyond $\sim600\,R_\odot$. Gas in the torus may either fall back onto the disk or become unbound, for instance through the release of recombination energy, which is not included in our simulations.
In Fig.~\ref{fig:keplerian} we show that the central CBD initially exhibits mostly sub-Keplerian rotation, as expected given its relatively large mass and high aspect ratio. In such a thick and massive disk, gas pressure contributes to the support of the disk. Once $v_{\rm kep}$ accounts for the mass of each central binary and the enclosed disk mass, the degree of rotational support is similar across mass ratios: $v_\phi/v_{\rm kep}$ is around 0.8--0.9 in the inner region close to the cavity and 0.4--0.6 in the main body of the disk. Comparing the runs with and without corotation ($q=0.5$), we find that corotation has only a small effect, with the run without corotation showing slightly lower values in the outer disk. The disk also moves towards Keplerian rotation over time as it cools and becomes thinner, with the ratio reaching $\sim0.8$--$1.0$ at $5\,000$ days for the run Q050-T100. The inner cavity, where the central binary resides, has a radius comparable to the orbital separation at the end of the plunge ($R_{\rm in} \sim a_f$; see $a_f$ in Table~\ref{tab:our_simulations}), and it grows to $\sim80\,R_\odot$ by $5\,000$ days in Q050-T100.

Taken together, these properties motivate our working definition of the CBD. We define the CBD as the bound gas that (i) is concentrated at the equatorial plane (Fig.~\ref{fig:histogram}); (ii) surrounds a low-density central cavity, showing that the bound portion of the envelope does not collapse onto the central binary (see Figs.~\ref{fig:temperature} and~\ref{fig:streamline}, where the cavity is especially visible at later times through its high temperature and low density in the center); (iii) is Rayleigh-stable (Fig.~\ref{fig:angular_momentum}), with rotation and pressure together providing the radial support against gravity; and (iv) is long-lived, persisting until the end of each simulation. We do not require the disk to be Keplerian: because of its large mass and high temperature, it is partially pressure-supported ($v_\phi/v_{\rm kep} < 1$, Fig.~\ref{fig:keplerian}).

The disk is therefore defined by two radii. Its inner edge, $R_{\rm in}$, is the edge of the central cavity, initially $\sim a_f$. Its outer edge, $R_{\rm out}$, is the radius at which the Rayleigh criterion first fails. At 500 days, $R_{\rm out} \approx 75$, 225, 285, 290, 340, and $490\,R_\odot$ for Q025, Q050, Q050-NCR, Q050-T100, Q075, and Q100, respectively (Fig.~\ref{fig:angular_momentum}). The outer edge increases over time; for Q050-T100, it grows from $\sim290\,R_\odot$ at 500 days to $\sim790\,R_\odot$ at $5\,000$ days. Although we do not simulate mass ratios below $q = 0.25$, we expect a minimum mass ratio below which a CBD does not form, since the angular momentum given to the envelope decreases with the companion mass and the Rayleigh stable region also decreases for smaller mass ratios.

\subsubsection{Radial Scale of Density}
\label{sec:density}

At a given location, the density of the disk can be approximated as

\begin{equation}\label{eq:density}
\rho(R,z) = \rho_0(R)\,\exp\!\left(-\frac{z^2}{2H(R)^2}\right),
\end{equation}

where $\rho_0(R)$ is the mid-plane density and $H(R)$ is the vertical scale height of the disk at radius $R$. Both quantities are estimated by fitting a Gaussian profile to the vertical density distribution within thin radial bins. The radial bins are constructed in logarithmic space, using 200 bins between $10$ and $1000\,R_{\odot}$. For each bin we compute the vertical density profile and determine $\rho_0(R)$ and $H(R)$ from the best-fitting Gaussian. The fact that the vertical density structure is well described by a Gaussian profile is consistent with a disk-like configuration. The resulting values of $H(R)$ and the corresponding aspect ratio $H/R$ are shown in Fig.~\ref{fig:height}.a and b, respectively. For the Q050-T100 run, $H(R)$ near the center is well fitted as:

\begin{equation}
    H(R) = [0.4(R/\text{R}_{\odot}) +10] \, \text{R}_{\odot}
\end{equation}

Beyond a certain radius, the quality of the Gaussian fit gradually decreases. This can be seen in Fig.~\ref{fig:height}.a, where $H$ departs from the linear fit. For the Q050-T100 run at 500 days, this departure becomes more pronounced at $\sim500~R_\odot$. Beyond this point, there is still ejected envelope material, but it is no longer well described by a Gaussian. In Fig.~\ref{fig:density_temperature}.a, we also show the radial profile of $\rho_0(R)$. Unlike the height profile, which follows a straight line, the density profile does not admit a simple fit. For the Q050-T100 run, a reasonable approximation is:

\begin{equation} \label{eq:central_density}
\begin{aligned}
\rho_0(R) =\;&
10^{-2}
\left(\frac{R}{10\,R_{\odot}}\right)^{-4} \\
&\times
\left[1+\left(\frac{125\,R_{\odot}}{R}\right)^{3.5}\right]^{-1.05}
\ \mathrm{g\,cm^{-3}} .
\end{aligned}
\end{equation}

For the other mass ratios, we did not attempt to derive an analogous analytical fit. Instead, we provide the density profiles shown in Fig.~\ref{fig:density_temperature}.a in the data files accompanying the paper. The functional form above reflects two regimes: a nearly flat inner region, close to the central cavity, and a steeper $R^{-4}$ decline in the main body of the disk. At larger radii, the material belongs to the outer, radially expanding component. This shell component carries away the mass of the envelope that is not in the disk at the beginning of our modeling. This is also consistent with the break found in the $H/R$ plots. This radially expanding shell is further discussed in Sec.~\ref{sec:shell}.

\subsubsection{Radial Scale of Temperature}
\label{sec:temperature}

\begin{figure*}
    \centering
     \includegraphics[width=0.94\linewidth]{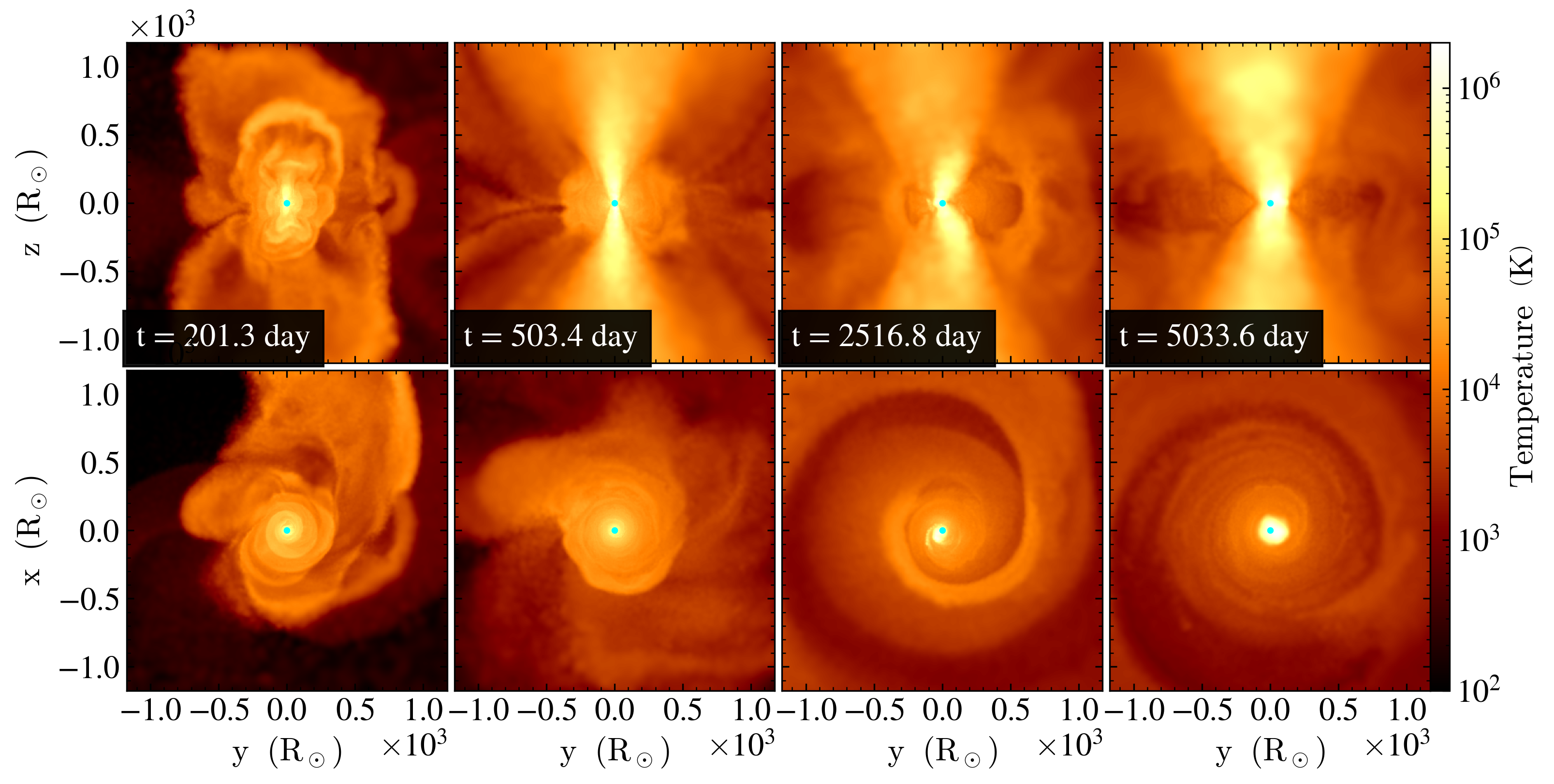}
    \caption{Temperature distribution for a $2~\text{M}_\odot$ RGB star in a binary system with mass ratio $q=0.5$, for a floor temperature of 100~K (Q050-T100 run).  Each panel shows both $y$-$x$ and $y$-$z$ planes at times $t = 201.3$, $503.4$, $2516.8$, and $5033.6$ days. The cyan '.' marks the center of mass of the core+companion binary.}
    \label{fig:temperature}
\end{figure*}

 Figure~\ref{fig:temperature} shows the temperature distribution in the $x$--$y$ and $y$--$z$ planes for the Q050-T100 case, in which we can see the region of the collimated outflows as well as the disk increasing its outer radius over time. We also examined the vertical temperature structure. Unlike the density, which follows a Gaussian profile, the temperature exhibits a more complex behavior in the $z$-direction, with a double peak near the mid-plane. To determine the radial scaling of the mid-plane temperature ($T_0$), we therefore consider a region with $|z| < 0.01R$ and compute the average temperature within this region.

In fig.~\ref{fig:density_temperature}.b, we show the radial profiles, which indicate that the mid-plane temperature follows a power-law behavior, $T(R) = A\times 10^4 \text{K}\,(R/100R_{\odot})^{-1}$. For the runs with a $10^4$~K floor temperature, we can see $T_0$ reaching a plateau at around $500$~R$_{\odot}$. Before this point, we did not see a difference between the runs with different floor temperatures. While the slope of the profile is consistent across the simulations, the normalization varies with the binary mass ratio. In our models the coefficient $A$ is approximately $4.5$ for the $q=0.5$ simulations (Q050 and Q050-T100). For the other runs, we provide data files of the plots.

\subsubsection{Mass of the disk}
\label{sec:disk_mass}

The mass of the circumbinary disk is obtained by integrating the disk
density profile over the full three-dimensional volume. Adopting the
Gaussian vertical structure of Eq.~(\ref{eq:density}),
$\rho(R,z) = \rho_0(R)\exp[-z^2/2H(R)^2]$, the mass is
\begin{eqnarray} \label{eq:mass_disk}
M_{\rm disk}
= \int_0^{2\pi}\int_{R_{\rm in}}^{R_{\rm out}}\int_{-\infty}^{\infty}
  \rho_0(R)\,e^{-z^2/2H(R)^2}\,R\,dz\,dR\,d\phi \nonumber
  \\
= 2\pi\sqrt{2\pi}\int_{\ln R_{\rm in}}^{\ln R_{\rm out}}
  \rho_0(R)\,H(R)\,R^2\,d\ln R ,\,\,
\end{eqnarray}
where $R_{\rm in}$ and $R_{\rm out}$ are the inner and outer radii of the CBD, as defined in Sec.~\ref{sec:disk_size}. At 500 days, the disk mass is $\sim0.57$--$0.78~M_\odot$ for the runs
with $q \geq 0.5$, and $\sim0.17~M_\odot$ for Q025.
For
comparison, the total bound mass, $(1-f_\text{unb})\,M_{1,\text{e}}$,
where $f_\text{unb}$ is the unbound fraction shown in
Fig.~\ref{fig:unbound_all} and $M_{1,\text{e}} = M_{1,\text{t}} -
M_{1,\text{c}}$ is the envelope mass, ranges from $\sim1~M_\odot$
(Q025) to $\sim1.3~M_\odot$ (Q100).
Thus, except for Q025, the disk
contains about half or more of the bound mass. In Q025, the disk holds
only $\sim17\%$ of the bound gas, consistent with its narrow Rayleigh-stable region
(Sec.~\ref{sec:disk_size}).

A substantial part of the remaining bound gas resides in a region between the outer edge of the disk and the radially expanding shell (Sec.~\ref{sec:shell}). For Q050-T100, extending the integration from $R_{\rm out} = 290~R_\odot$ to $500~R_\odot$ increases the enclosed mass from $\sim0.74~M_\odot$ to $\sim0.86~M_\odot$; that is, $\sim0.12~M_\odot$ of bound gas still follows a Gaussian vertical profile but is not rotationally supported.

\subsection{Torus and outer shell}
\label{sec:shell}

\begin{figure*}
    \centering
    \includegraphics[width=0.99\linewidth]{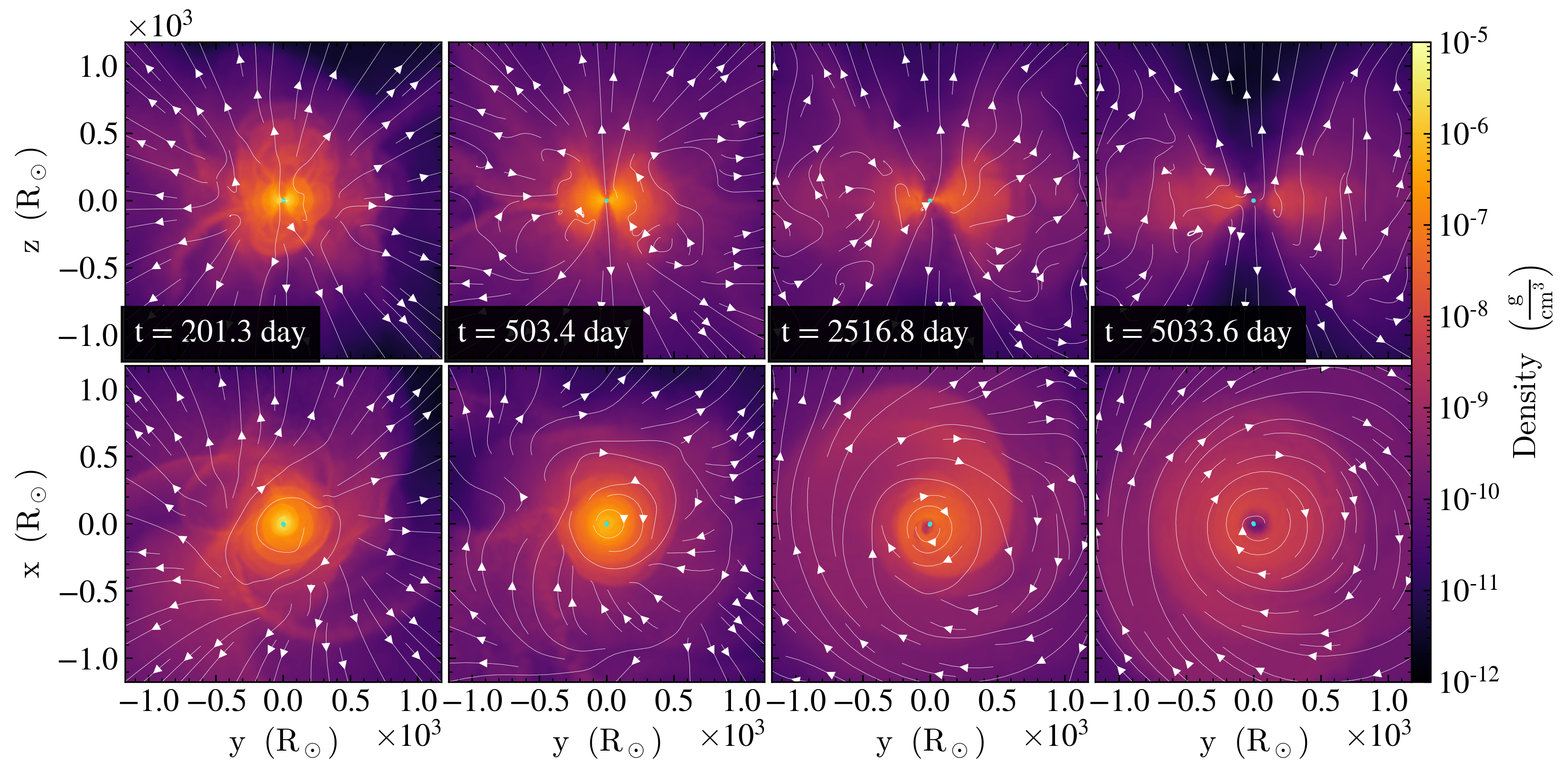}
    \caption{Density distribution for the same run~(Q050-T100), planes, and times as Fig.~\ref{fig:temperature}. White streamlines indicate the projected gas velocity field ($v_x$--$v_y$ in the $y$--$x$ plane and $v_y$--$v_z$ in the $y$--$z$ plane).}
    \label{fig:streamline}
\end{figure*}

\begin{figure*}
    \centering
     \includegraphics[width=0.99\linewidth]{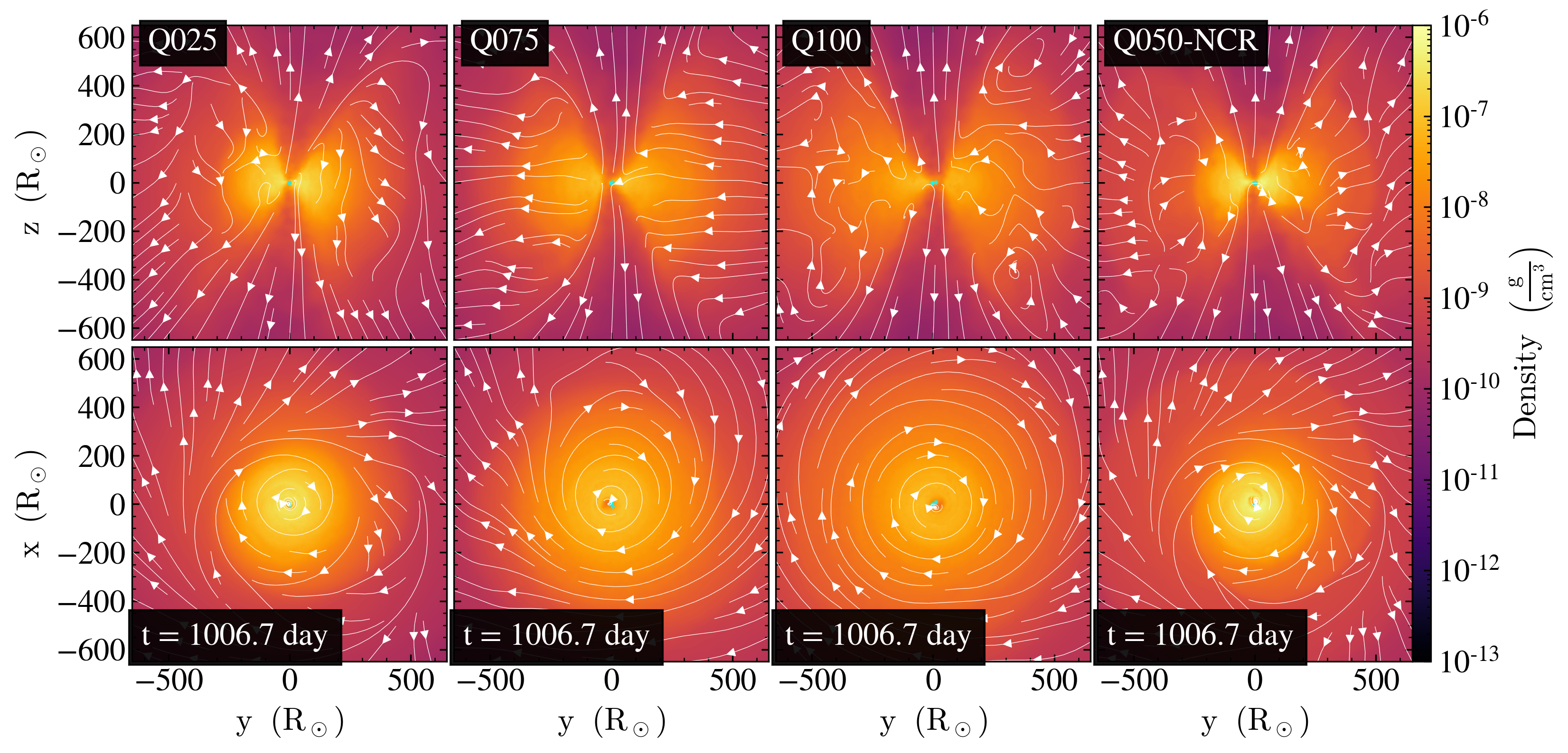}
    \includegraphics[width=0.99\linewidth]{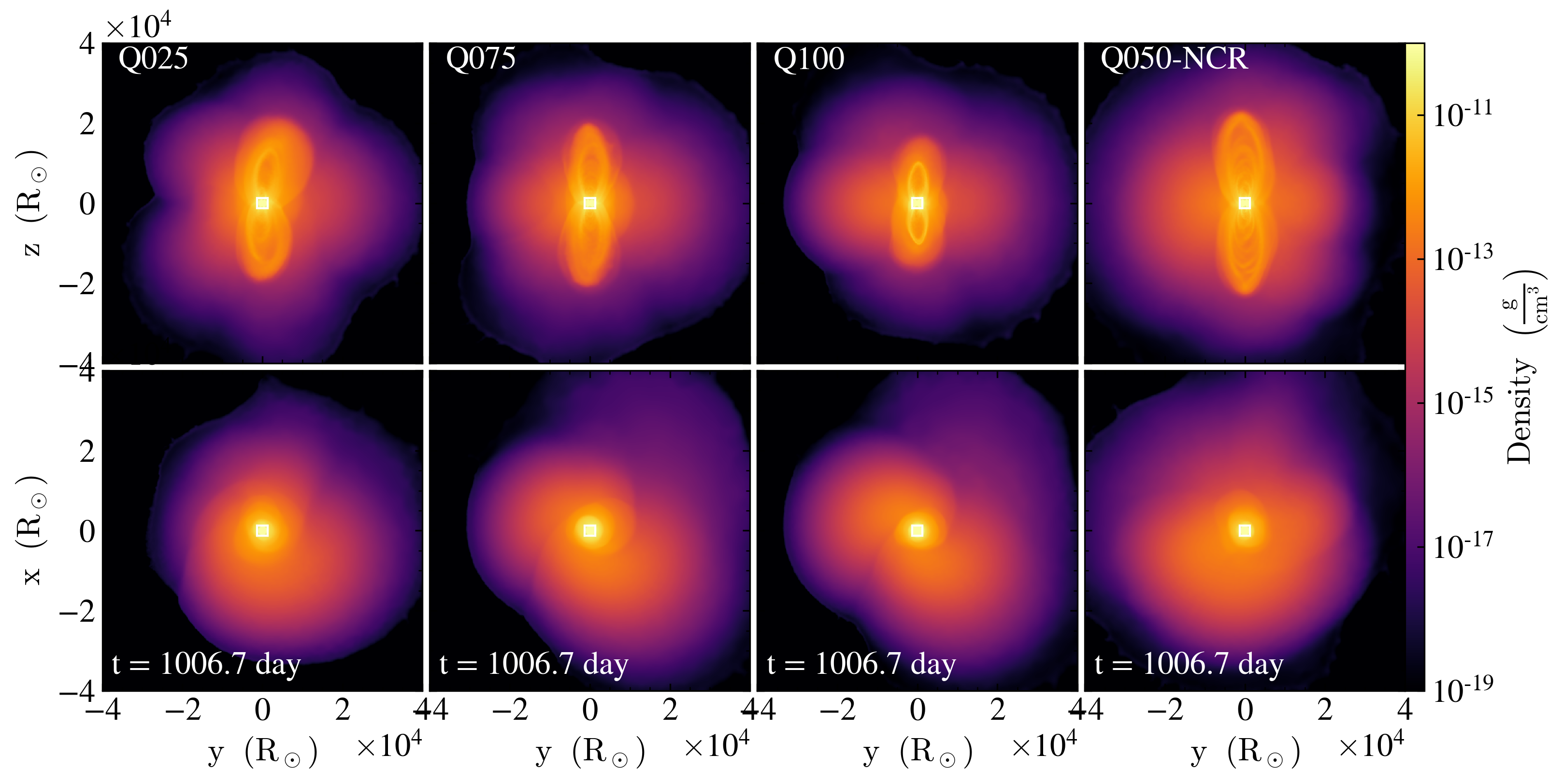}

    \caption{{\it Top:} Density distribution for the runs Q025, Q075, Q100, and, Q050-NCR at 1006.7 days. White streamlines indicate the projected gas velocity field ($v_x$--$v_y$ in the $y$--$x$ plane and $v_y$--$v_z$ in the $y$--$z$ plane). {\it Bottom:} Density distribution for the runs at a larger box. As we can see, in all runs, a bipolar shape is formed.}
    \label{fig:streamline_others}
\end{figure*}

Figs.~\ref{fig:streamline} and \ref{fig:streamline_others}~(Top) show the direction of the gas velocity streamlines for Q050-T100 at different times and for the other runs at $1\,006$ days, respectively. For Q050-T100 at 500 days, the flow is predominantly azimuthal in the $x$--$y$ plane for $R \lesssim 600\,R_\odot$ and develops a radial component at larger radii. In Fig.~\ref{fig:radial_velocity}.a we plot the radial velocity component $v_r$ as a function of radius. As shown, $v_r$ remains close to zero within the inner $\sim600\,R_\odot$ and then increases with radius as the flow becomes progressively more radial. In Fig.~\ref{fig:radial_velocity}.b we further examine how the kinetic energy is distributed among velocity components as a function of radius. At small radii, the kinetic energy is dominated by the azimuthal component, consistent with the presence of the disk. For Q050-T100 at 500 days, the radial component begins to dominate around $\sim600\,R_\odot$, and for $R \gtrsim 2000\,R_\odot$ the kinetic energy is entirely radial.

Outside the disk, therefore, the bound envelope does not immediately turn into free expansion. Between the outer edge of the CBD ($\sim290\,R_\odot$ for Q050-T100 at 500 days) and $\sim600\,R_\odot$ lies a transition region, which we refer to as a torus. The gas in this torus is still bound and has $v_r \approx 0$ (Fig.~\ref{fig:radial_velocity}), but it lies beyond the Rayleigh-stable region (Sec.~\ref{sec:disk_size}), so it is not rotationally supported and is not part of the disk. Nevertheless, it remains mostly concentrated toward the equatorial plane: its vertical density profile is still approximately Gaussian out to $\sim500\,R_\odot$ (Sec.~\ref{sec:density}), and between 290 and $500\,R_\odot$ it contains $\sim0.12\,M_\odot$ (Sec.~\ref{sec:disk_mass}). The torus is therefore an equatorial mass reservoir that is neither a disk nor part of the freely expanding shell, and its gas may either fall back onto the disk or become unbound. 

Beyond the torus, the envelope forms a radially expanding extended outer shell. A part of this shell is in homologous expansion ($v_r \propto R$). This outer layer is composed of envelope material ejected at the very beginning of the interaction, during the plunge. It corresponds to the $\sim10$--$20\%$ of the envelope that is unbound during the first $\sim200$ days, before the disk and the collimated outflows form (Fig.~\ref{fig:unbound_all}).

\begin{figure*}
    \centering
     \includegraphics[width=0.94\linewidth]{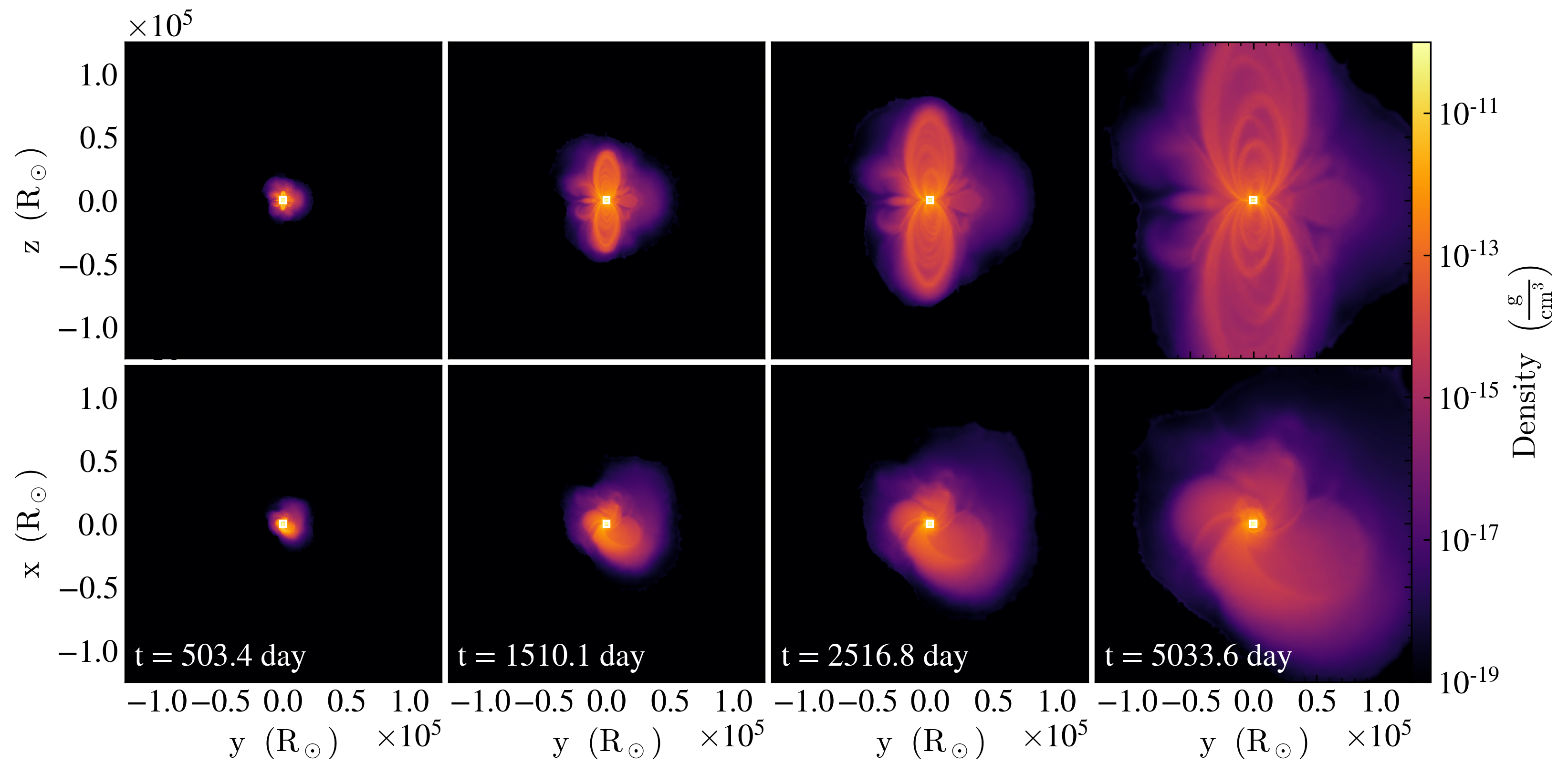}
    \caption{Density distribution at a larger box for the same run (Q050-T100), projections, and $t = 503.4$, $1510.1$, $2516.8$, and $5033.6$ days. The small white square at the center indicates the size of the entire region shown in Fig.~\ref{fig:streamline}, where the CBD resides.}
    \label{fig:zoomout}
\end{figure*}

\subsection{Collimated Outflows}
\label{sec:outflows}

\begin{figure}
    \centering
    \includegraphics[width=0.98\linewidth]{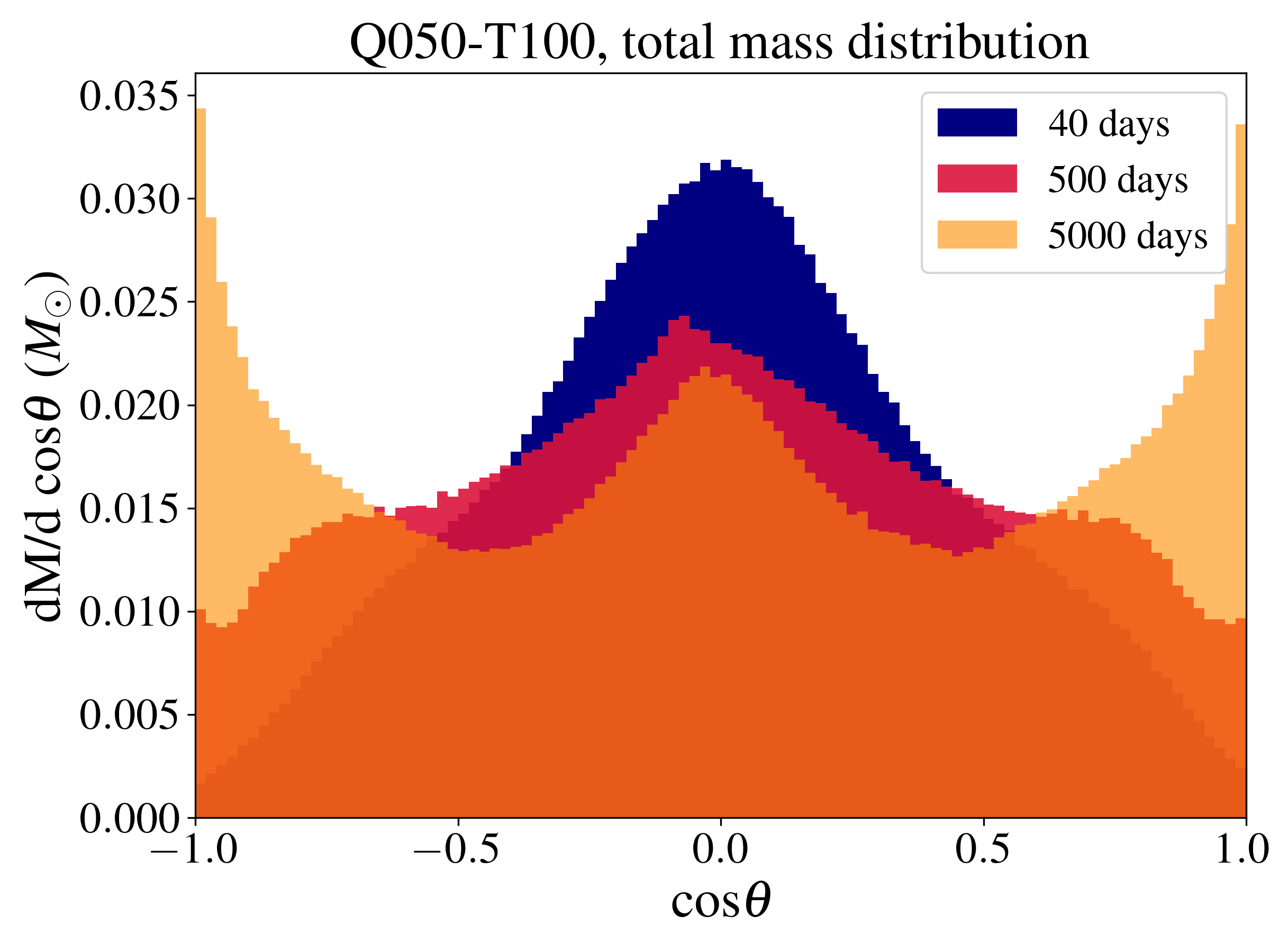}
    \caption{Mass distribution per $\cos(\theta)$ for the entire domain for the Q050-T100 run, at 40 (blue), 500 (red), and $5\,000$ (orange) days.}
    \label{fig:mass_distribution}
\end{figure}

From Figs.~\ref{fig:streamline_others}.b and~\ref{fig:zoomout}, we find that the bipolar structure of the outflow is clearly defined. The overall morphology is consistent with the bipolar structures commonly observed in PNe. This behavior is also reflected in Fig.~\ref{fig:mass_distribution}, which shows that the total mass contained in the polar regions increases with time, consistent with the preferential redistribution and ejection of envelope material toward the polar outflows. 

Our results are consistent with the mechanism proposed by \citet{2025A&A...697A..68G}, in which material falling from the CBD toward the central region is shock-heated and preferentially expands through the low-density polar directions. \citet{1992ApJ...389..628S} also proposed that bipolar outflows during CEE could arise from the transfer of angular momentum to the envelope. 

\subsubsection{Opening angle of the outflow}

\begin{figure*}
\includegraphics[width=0.47\textwidth]{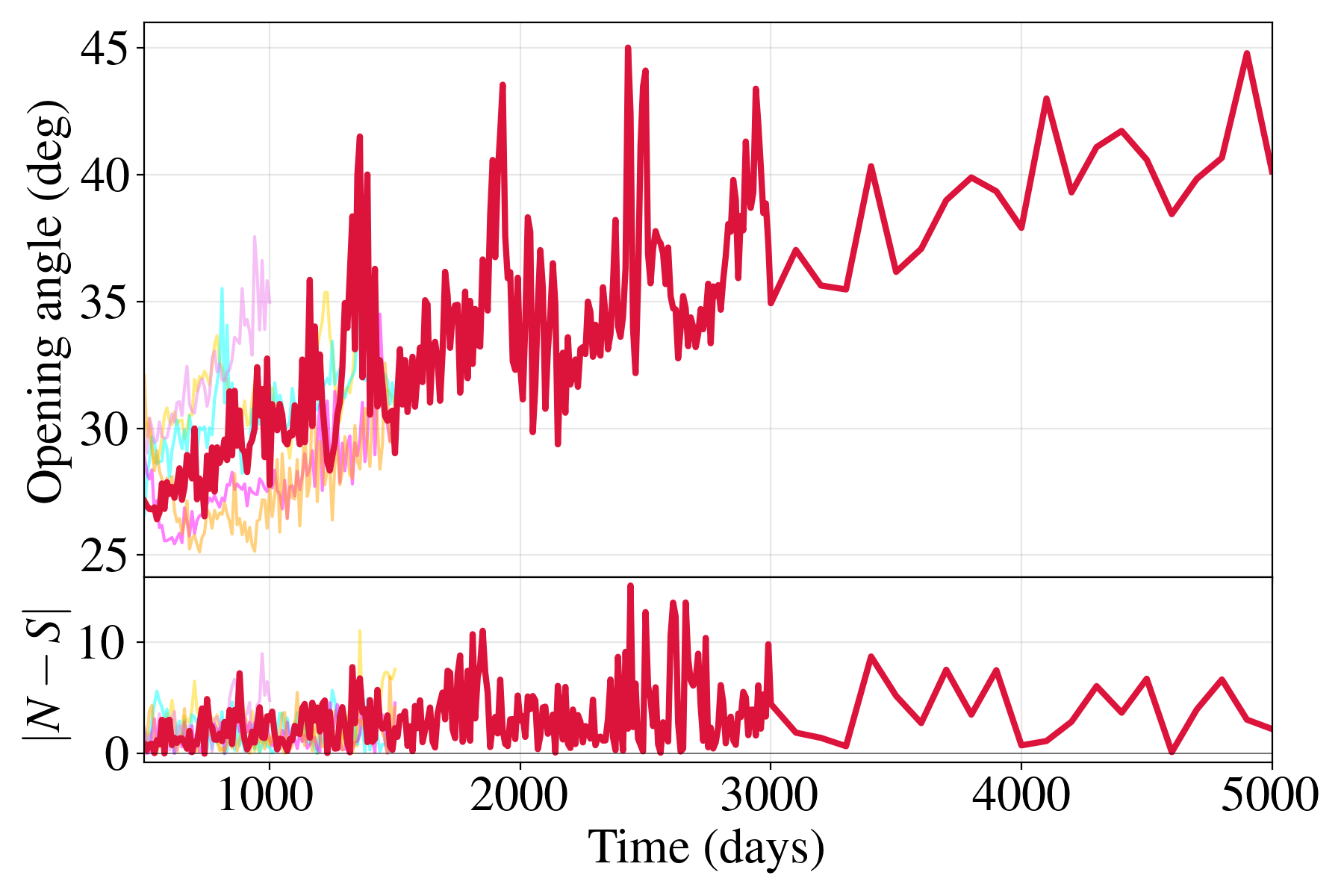}
\includegraphics[width=0.47\textwidth]{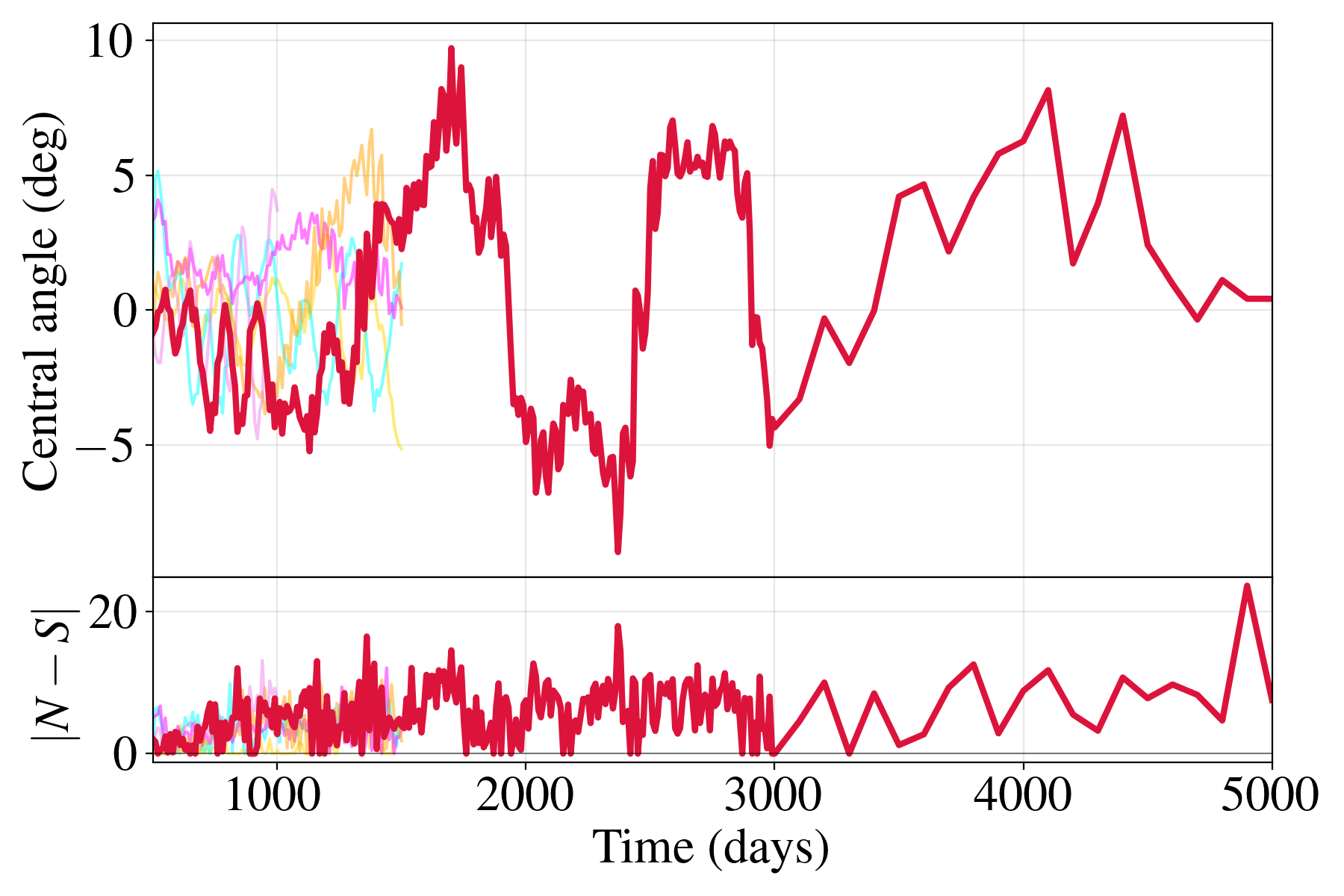}
\includegraphics[width=0.47\textwidth]{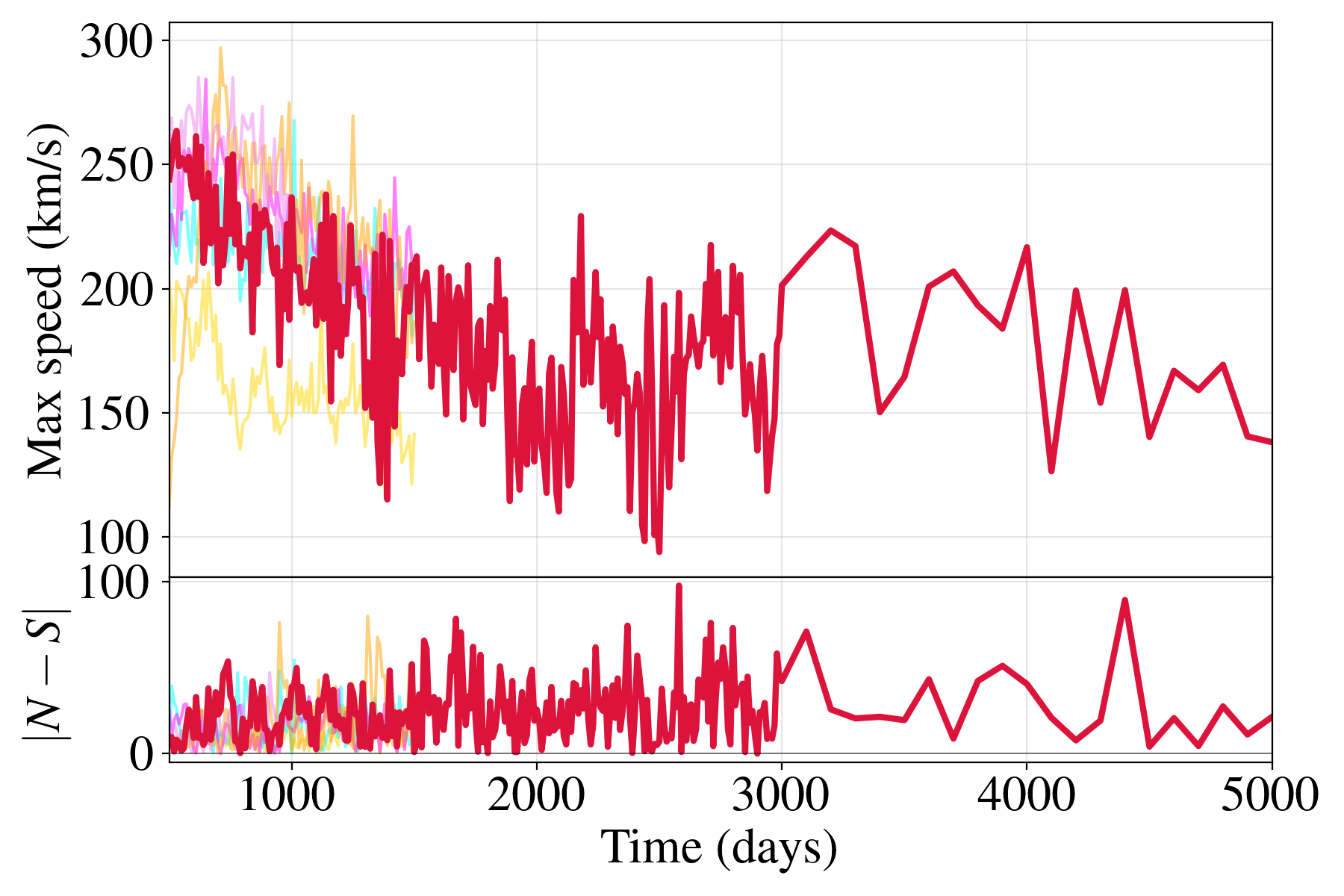}
\includegraphics[width=0.47\textwidth]{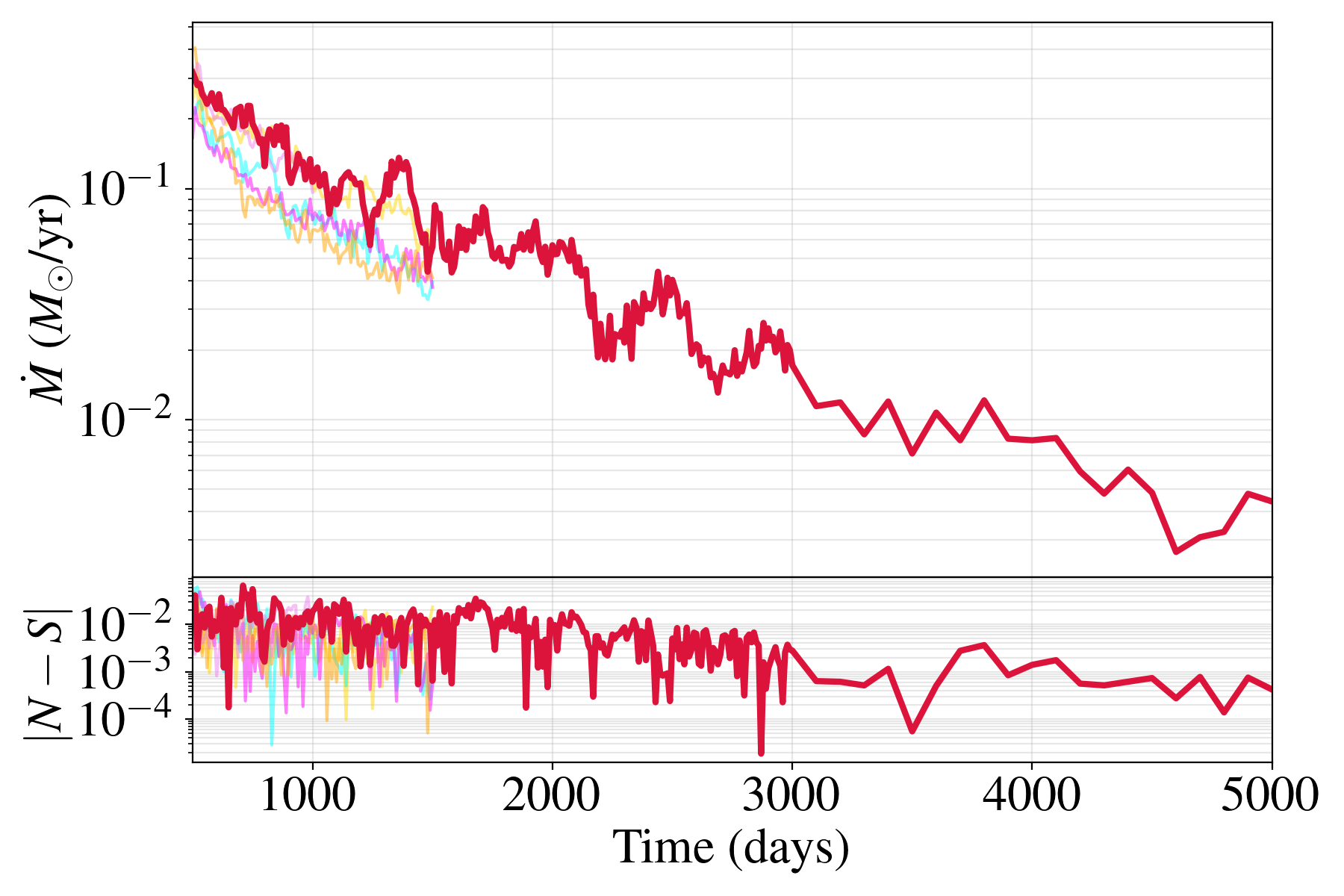}
\caption{{\it Top Left:} Opening angle versus time. {\it Top Right:} Central angle versus time. {\it Bottom Left:}  Maximum speed versus time. {\it Bottom Right:} Mass-loss rate through the poles versus time. The colors per simulation run are the same as Fig~\ref{fig:unbound_all}. For the run Q050-T100 the final 2000 days have less temporal resolution, having one output every 100 days, as opposed to every 10 days for the rest of the simulations. Each quantity is computed separately for the northern ($z>0$) and southern ($z<0$) outflows. The main panels show the mean of the two hemispheres (the sum, for $\dot{M}$), and the lower sub-panels show the absolute difference between them, $|N-S|$, as a measure of the north–south asymmetry of the outflows.}

\label{fig:time_jet}
\end{figure*}

In Fig.~\ref{fig:time_jet}.a, we show the evolution of the opening angle of the outflow as a function of time for all simulations using the \texttt{HLLC} solver and 1.2 million star particles. To estimate the opening angle, we select a spherical shell with a radius between $290$ and $300~R_\odot$, located near the base of the outflows. Depending on the mass ratio, this shell lies inside (Q075, Q100), near (Q050-T100), or outside (Q025, Q050) the Rayleigh-stable disk (Sec.~\ref{sec:disk_size}); nevertheless, a low-density polar funnel bounded by flattened equatorial material is present at this radius in all runs (Figs.~\ref{fig:gaussian} and~\ref{fig:streamline_others}), so the opening angle traces the geometry of the funnel rather than that of the rotationally supported disk itself. The velocity field within this shell is binned into 100 angular segments, and the resulting distribution is fitted with a Gaussian profile. The opening angle is then defined as $2\sigma$ of the Gaussian distribution (see Fig.~\ref{fig:gaussian} for an illustration of the Gaussian fit and the circle at $300~R_\odot$). We perform this fit independently for the northern and southern outflows; the difference between the two, $|N-S|$, shown in the lower sub-panels of Fig.~\ref{fig:time_jet}, quantifies how asymmetric the bipolar outflows are.

We find that the opening angle increases with time, as expected from the spreading of the disk and also the increasing amount of ejected material, which reduces the pressure near the poles. In the Q050-T100 run, the opening angle increases from $\sim30^\circ$ to $\sim40^\circ$ over the $5\,000$ days of the simulation. We also find some dependence on the mass ratio, with different mass ratios producing slightly different opening angles, although the difference is less than $10^\circ$. The central angle~(the peak position of the Gaussian fit or, in other words, how much the outflow axis is misaligned with the $z$--axis) varies between $0^\circ$ and~$10^\circ$ over time in all runs. Because of that, the time-averaged apparent opening angle is larger.

\subsubsection{Speed of the outflow}

The maximum speed of the ejecta ($v_{\rm max}$ at $\theta = 0^\circ$) begins at approximately $250~\mathrm{km\,s^{-1}}$ and gradually decreases to about $200~\mathrm{km\,s^{-1}}$ during the evolution. These velocities are comparable to the escape speed from the central binary at its orbital separation, $v_{\rm esc} = [2G(M_{1,\rm c}+M_2)/a_f]^{1/2}$, which ranges from $\sim230$ to $\sim310~\mathrm{km\,s^{-1}}$ across our runs. At a distance of $300~R_\odot$ from the binary, however, the escape speed in all simulations is below $100~\mathrm{km\,s^{-1}}$, which is smaller than the velocity corresponding to $1\sigma$ of the Gaussian distribution ($0.606\,v_{\rm max}$). This indicates that the out-flowing material exceeds the local escape speed and is therefore able to become unbound from the central binary. For the Q050-T100 run, at later times ($5\,000$ days), the outflow velocity has decreased substantially to around $150~\mathrm{km\,s^{-1}}$.

A comparable dependence on the system parameters was reported by \citet{2022A&A...660L...8O}, where the maximum velocity varied between $\sim120~\mathrm{km\,s^{-1}}$ and $60~\mathrm{km\,s^{-1}}$ depending on the mass ratio. In their case, the final separation of the simulations is larger, which explains the smaller escape speeds and, consequently, the lower outflow velocities. In general, these studies indicate that the characteristic velocity of the outflow is of the same order as the escape speed of the central binary. Another interesting property is that the maximum velocity oscillates with time. These variations in the outflow velocity can produce higher density regions given that faster material catches up with slower material ejected at earlier times, naturally producing the multiple shells visible within the bipolar outflows in Fig.~\ref{fig:zoomout}.

\subsection{Mass-loss rate}
\label{sec:mdot}

For the run Q050-T100, the mass-loss rate is $\sim 3\times 10^{-1}\,M_{\odot}$~yr$^{-1}$ at early times, decreasing following a power-law behavior thereafter, reaching $\sim 6\times10^{-3}\,M_{\odot}$~yr$^{-1}$ at $t=5\,000$~days. This was estimated by computing the average value of $\rho v_r$ within each angular bin $\theta_i$ on a spherical shell, multiplying by the area of the corresponding polar cap, and summing over all bins within the collimated outflow region, defined as the region within an angle $\sigma$ of each pole, where $\sigma$ is the width of the Gaussian at each time, as shown in Fig.~\ref{fig:gaussian}.

\begin{figure*}
\includegraphics[width=0.9\textwidth]{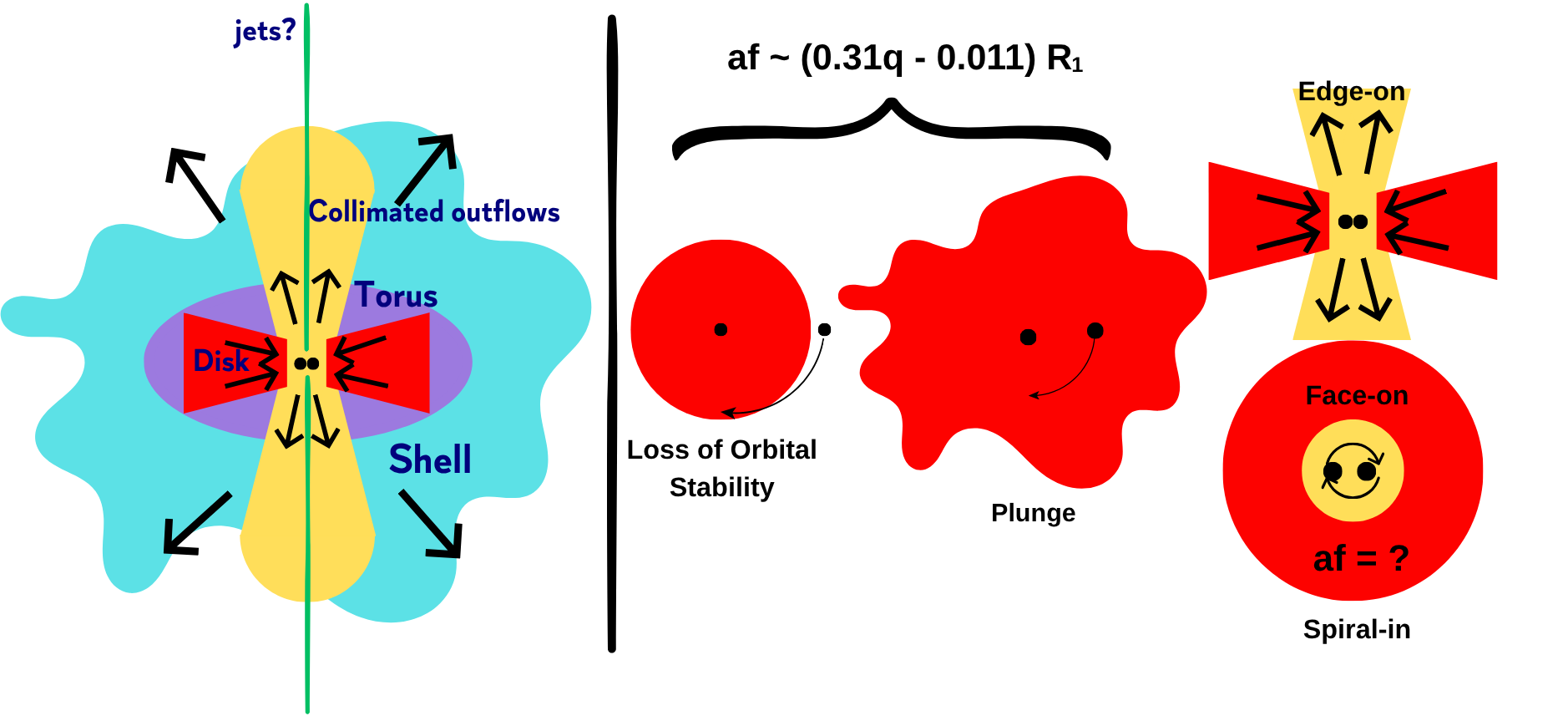}
\caption{{\it Left:} Simplified schematic representation of the different components of the ejecta for a classical CEE right after the plunge phase. The components found in this work are the outer shell, the torus, the CBD, and the collimated outflows. We also include the narrower jets seen in several observations, which are not captured by our simulations. Because the fate of the torus is uncertain (it may fall back onto the disk or become unbound), we do not draw arrows for it. {\it Right:} Simplified description of the three different phases of a classical CEE. The final separation at the end of the plunge phase is taken from our previous work, \citet{2026ApJ...998...34B}.}
\label{fig:multi_component}
\end{figure*}

\section{Comparing our simulations with observation}
\label{sec:observations}
Comparison between observations and simulations of CEE is challenging because no single class of objects is ideal: each offers a different trade-off between timing, i.e., how close the observed epoch is to the phases we simulate, and certainty, i.e., how confidently the object can be identified as the product of CEE. The relevant classes, LRNe, post-AGB stars (in particular WFs), and PNe, should not, however, be viewed as independent. They likely probe different stages of the same process, and a subset of systems in one class may evolve into another. Considering them together therefore allows us to combine the strengths of each: the earlier classes have the right timing to be compared to our simulations, while the later ones are more securely identified as CEE products.

LRNe are the closest in time to our simulations, since they are expected to trace the plunge itself. As discussed in the introduction, most LRNe are regarded as stellar mergers, but a transient event similar to, and perhaps indistinguishable from, a "merger LRN" is also expected for successful CEE~\citep[e.g.,][]{2013Sci...339..433I,2026ApJ...999..217C}. No such event has yet been conclusively identified, because the surrounding gas and newly formed dust hide the central system, and a newly formed close binary cannot yet be told apart from a merger remnant. Moreover, with a few exceptions~\citep[e.g., AT~2025abao;][]{2026A&A...709A..25R}, the properties of the progenitor systems are unknown. Although most extragalactic LRNe arise from massive progenitors, the Galactic LRNe, for which ALMA has provided most of the available morphological information~\citep{2018A&A...617A.129K,2024A&A...682A.127S}, have low- to intermediate-mass progenitors, closer to the systems simulated here. Even if these events are mergers, some of their properties should be transferable to classical CEE. The CBD, for instance, forms from the angular momentum transferred from the orbit to the envelope during the plunge, a process that happens whether or not the binary ultimately merges.

For the spiral-in phase of CEE, there is one promising class of observations: Water Fountains (WFs). Most of them are post-AGB stars, although the class spans the end of the AGB, as in W43A~\citep{2020ApJ...890L..14T}, to the beginning of the PN phase, as in IRAS~15103$-$5754~\citep{2018MNRAS.480.4991G}. WFs have dynamical ages of at most $\sim 200$ years and are identified by their H$_2$O masers, which trace fast bipolar outflows with velocities of $20$--$800~\mathrm{km~s^{-1}}$, and $\sim$~18 sources are currently confirmed~\citep{2025A&A...703A.268C}. WFs also host an equatorial torus containing a substantial fraction of the original envelope mass and expanding at a lower velocity. WFs are still very young and also suffer from the obscuration problem of LRNe, and thus no close binary has been directly detected in WFs. Nonetheless, their extreme bipolarity has long been linked to CEE. More recently, \citet{2022NatAs...6..275K} presented several arguments, beyond morphology, supporting a CEE origin for these objects. The authors noted that WFs were previously associated with intermediate-mass ($>5\,\text{M}_\odot$) post-AGB stars because of their high mass-loss rates ($\sim 10^{-4}$--$10^{-3}\,\text{M}_\odot\,\mathrm{yr^{-1}}$). However, their formation rate and $^{18}$O abundances are inconsistent with this interpretation. Moreover, the oxygen-rich chemistry of WFs indicates that their envelopes were ejected before the third dredge-up could enrich the envelope with carbon. Overall, the authors argue that CEE is the only known event capable of explaining all of these observations. Because WFs are still ejecting a substantial part of their envelope, they may be providing an observational window of the elusive spiral-in phase of CEE.

PNe lie at the opposite end: they are the furthest in time from our simulations, but the most securely linked to CEE, since close binary central stars have been identified in a number of them~\citep[see, for a list of some PNe harboring close binaries,][]{2019ibfe.book.....B}. PNe are shells of ionized gas and dust produced by giants at the end of their lives~\citep{2022PASP..134b2001K}. Approximately 80\% of PNe exhibit asymmetric morphologies, and about 22\% display bipolar structures~\citep{1998AJ....116.1357S,2006MNRAS.373...79P,2021A&A...656A..51G}, which are difficult to reproduce through single-star evolution but are expected naturally from CEE. Moreover, because a PN is a few thousand years old, it can preserve a fossil record of the entire ejection history, potentially including material ejected during the loss-of-orbital-stability, plunge, and spiral-in phases, as well as material shaped by later low-density winds from the pre-white-dwarf core~\citep[e.g.,][]{2018ApJ...860...19G,2020MNRAS.497.2855Z}.

Each of these classes carries its own uncertainties, and any comparison is therefore on an order-of-magnitude basis. Nonetheless, some of the same properties emerge across all of these classes: LRNe, WFs, post-AGB stars, and post-CEE PNe show collimated outflows and equatorial structures consistent with a disk. This is consistent with the bipolar morphology and the CBD being present from the end of the plunge, as in our simulations. Our goal is thus to assess whether the overall morphology emerging from our simulations is consistent with these observations.

\subsection{CBD and Torus}
\label{sec:cbd_obs}

 Evidence for disks extends back to LRN remnants, which are consistent with our simulations in terms of timeline. Infrared excesses in the spectral energy distributions of V4332~Sgr and V1309~Sco have been interpreted as disks~\citep{2010A&A...522A..75K,2015A&A...580A..34K}. Observations of WFs also indicate massive equatorial structures similar to what is expected for a CBD or torus, which is consistent with the picture that the ejecta from CEE is not ejected all at once during the plunge. For example, IRAS~15103$-$5754 and IRAS~16342$-$3814 are estimated to host equatorial tori with masses of $0.4$--$1.0\,M_{\odot}$ and $\sim 1\,M_{\odot}$, respectively~\citep{2018MNRAS.480.4991G,2012A&A...544A..58M}. These masses are comparable to those of our CBDs + torus at 500 days ($\sim0.6$--$0.8\,M_\odot$ for $q \geq 0.5$). One caveat applies to this comparison: the progenitor masses of these WFs are unknown. Objects from other classes related to CEE show similar features: CK~Vul, a CEE-merger remnant, very likely hosts a disk~\citep{2020A&A...644A..59K}, and the Boomerang Nebula, a post-RGB system associated with CEE, hosts a dusty flared disk~\citep{2013ApJ...777...92S,2017ApJ...841..110S}.

From our simulations, we expect the CBD to be thick and possibly massive shortly after its formation, consistent with the massive tori of WFs, and to become thinner over time (in Fig.~\ref{fig:height} we can see the thickness at $5\,000$~days being smaller than at 500~days). Thus, for PNe we expect the CBD to have become thinner, more extended, but still present. Several post-CEE PNe, such as the Red Spider Nebula~\citep{2025ApJ...993...79K} and the Necklace Nebula~\citep{2026A&A...707A.169J}, host equatorial disk-like or ring-like structures, so equatorial structures persist into the PN phase, at least for a subset of objects. For objects older than PNe, close binaries containing hot subdwarfs, which point to a CEE during the RGB phase, as in our simulations, also show infrared excesses interpreted as CBDs~\citep{2025MNRAS.537.1950L}.

There is also evidence, albeit controversial, of disks leaving imprints in even older systems. Second-generation planets formed from a CEE disk rather than a protoplanetary disk have also been proposed in some systems, such as NN~Serpentis~\citep{2013A&A...555A.133B}, although both their detection~\citep{2023MNRAS.526.4725O} and formation channel~\citep{2025PASA...42..159P} are not settled. If confirmed, this might suggest that some post-CEE disks are massive enough to become gravitationally unstable. Moreover, some polars show mid-infrared emission consistent with a long-lived CBD, although an alternative origin for that emission has been debated~\citep{2006ApJ...646L..65H}. 

Taken together, at least for CEE with low-mass giants, CBDs appear to be widespread and to harbor a substantial fraction of the original envelope mass, in agreement with our simulations. Observations consistent with CBDs have been reported across a diverse set of CEE-related systems: post-RGB systems such as the Boomerang Nebula and hot subdwarf binaries, post-AGB systems such as WFs, and CEE merger remnants such as LRNe and CK~Vul. Although how common CBDs are remains to be quantified, the present work offers a natural formation pathway for such structures and shows that they can form across a range of mass ratios (at least $0.25\leq q \leq1$, according to our work).

\subsection{Collimated outflows}
\label{sec:outflows_obs}
In our simulations, the bipolar outflows form early, during the transition from the plunge to the spiral-in, and remain collimated while propagating through the earlier-ejected envelope out to $\sim10^5~R_\odot$ in $\sim 14$~years~(Fig.~\ref{fig:zoomout}). This addresses the concern raised by \citet{2025RAA....25b5023S} that hydrodynamical outflows are too weak and would be deflected before forming large-scale bipolar lobes. An early bipolar morphology is consistent with observations of LRNe, such as AT~2025abao~\citep{2026A&A...709A..25R}, and with ALMA observations of the Galactic LRN remnants V4332~Sgr, V838~Mon, and V1309~Sco~\citep{2018A&A...617A.129K,2023atyp.confE...7S,2024A&A...682A.127S}. It also implies that the bipolar geometry of PNe does not necessarily require late-time winds~\citep[cf.][]{2018ApJ...860...19G,2020MNRAS.497.2855Z}: the geometry in Fig.~\ref{fig:zoomout} is similar to the wider bipolar component of the Red Spider Nebula~\citep{2025ApJ...993...79K}. Winds and jets are nonetheless expected to modify this shape later. For instance, NGC~2818 (dynamical age of the lobes of $7\,400$--$11\,000$~yr) still has a torus-like and bipolar structure, but has a more deformed morphology~\citep{2026ApJ...998....7M} than the younger Red Spider~\citep[dynamical age of the lobes of $3\,700$~yr;][]{2025ApJ...993...79K}.

The opening angles of our outflows are also consistent with observations. For V4332~Sgr, the opening angle of the bipolar component is $\sim60^\circ$~\citep{2018A&A...617A.129K}, larger than, but of the same order as, the values we find ($\sim30$--$45^\circ$). Our opening angle is defined as $2\sigma$ of the Gaussian fit to the velocity distribution near the base of the outflow, whereas observational estimates trace the geometry of the lobes, so we expect the two measures to be of similar magnitude but not identical. In addition, the outflow axis in our simulations wobbles by up to $\sim10^\circ$ over time, which broadens the time-averaged apparent opening angle. Similar values are found for the wide-angle component of WFs: W43A, IRAS~15103$-$5754, and IRAS~18043$-$2116 have opening angles of $40^\circ$, $56^\circ$, and $60^\circ$, respectively~\citep{2025ApJ...987..114I,2018MNRAS.480.4991G,2023ApJ...948...17U}. In contrast, the magnetically driven outflows of \citet{2025A&A...698A.133V} are narrower, with opening angles of $16$--$24^\circ$, suggesting that the wide-angle component of WFs is better reproduced by hydrodynamical collimation by the disk, while magnetic fields may be more relevant for the narrower jets.

The speed and mass-loss rate of the outflows are harder to compare. We expect the ejecta speed to be of the same order as the escape speed of the central binary, and thus vary substantially from system to system. Moreover, because the binary remains obscured during the LRN and WF phases, and for LRNe it is often unclear whether a merger has happened, the escape speed of the central system is usually unknown. Nonetheless, ejecta speeds of $\sim200~\mathrm{km\,s^{-1}}$ are found both for V4332~Sgr and for some WFs~\citep{2025A&A...703A.268C,2025A&A...703L...7C}, comparable to those in our simulations. The mass-loss rates of WFs, $\sim10^{-4}$--$10^{-3}\,M_\odot\,\mathrm{yr^{-1}}$~\citep{2022NatAs...6..275K}, depend on the system parameters and on physics not included in this work, such as cooling and accretion. Still, extrapolating the mass-loss rate of Q050-T100, which declines approximately as $\dot{M}\propto t^{-2}$ over the first $\sim14$~yr, to $\sim100$~yr gives $\dot{M}\sim10^{-4}\,M_\odot\,\mathrm{yr^{-1}}$, broadly comparable to the observed rates, although this extrapolation neglects the missing physics.

\subsection{Outer shell}

Most post-CEE objects are surrounded by material lost before the CEE, either through AGB winds or through earlier binary interactions. As the shell expands into this pre-existing material, it is decelerated and shock-heated, so its original velocity, temperature, and morphology are quickly erased, making it hard to disentangle the CEE ejecta from the surrounding medium. Post-RGB systems offer a cleaner view: because mass loss from single-star evolution on the RGB is negligible, in the absence of earlier binary interactions the shell can expand nearly freely and preserve its original properties.

One such candidate is the Boomerang Nebula, proposed to be the product of a CEE that began while the primary was an RGB or early-AGB star and ended in a merger~\citep{2017ApJ...841..110S}. As discussed in Sec.~\ref{sec:cbd_obs}, its central region hosts a dusty, flared disk, and the system shows collimated bipolar outflows~\citep{2013ApJ...777...92S, 2017ApJ...841..110S}. Surrounding these is an ultra-cold ($\sim 1$\,K), roughly spherical shell extending to $\gtrsim 1.2\times10^{5}$\,au, expanding at $\sim 164\,\mathrm{km\,s^{-1}}$ with a velocity that increases linearly with radius, and with a mass of $\gtrsim 3.3\,M_\odot$~\citep{2017ApJ...841..110S}. Its temperature, below that of the cosmic microwave background, is attributed to the adiabatic expansion of fast gas into a region largely free of earlier circumstellar material. This is the expected behavior of the outer shell in our simulations, which is ejected during the plunge, before the disk forms, and approaches homologous expansion. We cannot follow this cooling down to $\sim 1$\,K, which is well below our temperature floor. Moreover, the Boomerang's progenitor ($\gtrsim 4\,M_\odot$ on the main sequence;~\citealt{2017ApJ...841..110S}) is more massive than ours, and a much larger fraction of its envelope resides in the fast component, so a direct quantitative comparison is not possible. Nonetheless, its homologous expansion and velocity are consistent with an outer shell ejected during the plunge, like the one found in our simulations.

\section{Discussions and Conclusions}\label{sec:discussion}

We performed ten 3D hydrodynamical simulations of CEE with a $2~\mathrm{M}_\odot$ RGB donor and mass ratios $0.25 \leq q \leq 1$, following the system from the plunge into the early spiral-in. In all our fiducial (\texttt{HLLC}) runs, the ejecta develop four components: an outer shell component, carrying the $\sim10$--$20\%$ of the envelope unbound during the plunge; a thick, massive CBD that is sub-Keplerian and Rayleigh-stable; a torus of bound equatorial gas between the disk and the shell, which is not rotationally supported and at later times may either fall back onto the disk or become unbound; and wide-angle bipolar outflows, with opening angles of $\sim30$--$45^\circ$ and speeds of $\sim150$--$250~\mathrm{km\,s^{-1}}$. The disk and outflows form between 150 and 300~days, at $\epsilon \sim 10^{-3}$, later than the commonly adopted threshold $\epsilon = 0.01$ for the end of the plunge phase. 

These components are summarized schematically in Fig.~\ref{fig:multi_component}, which is inspired by the multi-component outflows of protostars~\citep{2025A&A...699A.361V}, with which CEE ejecta share morphological similarities. They can be regarded as the minimum set that the morphology of a classical CEE should have. Real systems are likely more complex still. First, CEE can leave imprints of earlier mass-loss events that remain recorded in the resulting PN. The PN Fleming~1 (Fg~1) illustrates this well: it shows two distinct mass-loss events, an older one, traced by jet-like structures $\sim2\,500$--$7\,000$~yr old and suggestive of an earlier, possibly symbiotic-like, mass-transfer phase, and a younger one, $\sim2\,000$~yr old, associated with the CEE event~\citep{1993A&A...267..194L,2012Sci...338..773B,2019ibfe.book.....B}. 

Second, because we include neither accretion nor magnetic fields, our simulations do not self-consistently form jets, which are present in some post-CEE PNe and WFs (Sec.~\ref{sec:outflows_obs}). Our outflows arise from pure hydrodynamics, without magnetic fields, but their formation depends on the Riemann solver: they form with \texttt{HLLC} but not with the more diffusive \texttt{HLL}. Although magnetic fields are not necessary for the formation of these outflows, they have been detected in some WFs~\citep{2006Natur.440...58V,2013MNRAS.436L..79P,2015ApJ...806..105S} and are expected to influence the evolution, such as the final separation and local asymmetries of the outflows~\citep{2025A&A...698A.133V}. Magnetic fields may therefore modify the structure, velocity distribution, and variability of the outflows, but the large-scale wide-angle bipolar morphology arises from purely geometrical constraints in our simulations. Third, marginally bound material in the torus may fall back onto the disk, adding further structure over time. Finally, recombination likely affects real systems, but to an extent that remains debated: \citet{2018MNRAS.478.1818G} argued that much of the recombination energy is radiated away and contributes little to envelope removal, whereas simulations that include it find recombination-driven outflows, which would constitute an additional component of the ejecta~\citep{2024A&A...691A.244V}.
 
The picture becomes even more complex once we consider that not all CEEs are classical. Broadly, we can distinguish three main outcomes: (1) a close binary is formed, likely surrounded by a short- or long-lived CBD; (2) the core and companion merge, but the energy and angular momentum deposited in the envelope are sufficient to form a disk and subsequently unbind the envelope, as in CK~Vul~\citep{2020ApJ...904L..23B}; and (3) the core and companion merge, but the energy and angular momentum are insufficient to form a disk, so most of the envelope remains bound to the merger product and only part of it is ejected. Examples of the latter are the proposed merger origins of the progenitor of SN~1987A~\citep{1989A&A...219L...3H} and of $\eta$~Carinae~\citep{2016MNRAS.456.3401P}. Moreover, a merger can happen during the plunge or later, during the longer spiral-in, in which case the interaction between the central binary and the surrounding envelope is essential for this "delayed" merger. This highlights the importance of understanding whether and how disk formation depends on the mass ratio, giant mass, and evolutionary stage, in order to predict whether a "delayed" merger happens.

CEE has remained one of the major unsolved problems in stellar astrophysics for the past fifty years, in part because, as discussed above, its outcome differs from case to case. Most often, however, "solving" CEE refers to predicting the final separation, i.e., case (1) above, and even in this case we lack both a predictive model and a clear consensus on how each phase of the evolution proceeds. Here we found that, when a CBD forms, the last phase of CEE, the spiral-in, should be treated as a binary surrounded by a CBD. Initially, the disk is geometrically thick, so classical CBD theory should be applied with caution. The impact of a thin CBD on the orbital separation of a central binary has been widely studied, and such a disk can, in principle, either widen or shrink the orbit~\citep{2023ARA&A..61..517L,2024ApJ...970..156D}. \citet{2020ApJ...900...43T} showed that the outcome can depend on disk thickness, but their study explored only aspect ratios in the range $H/R = 0.01$--$0.1$. The CBD produced immediately after the CEE plunge has $H/R \sim 0.3$--$1$~\citep[see Fig.~\ref{fig:height}; also][]{2024A&A...691A.244V}, a regime that, to our knowledge, remains unexplored in studies of binary--disk interaction. We believe that future studies of the evolution of thick CBDs like those formed here will provide valuable insight into the final phase of CEE, even without modeling the full process self-consistently.

In Fig.~\ref{fig:multi_component}, we show a schematic description of the different phases of classical CEE. By the end of the plunge, the final separation can be estimated as $a_f \sim (0.31q-0.011)R_1$, a result from our previous work~\citep{2026ApJ...998...34B}, while the morphology of the ejected envelope is presented in this work. Key open questions we expect to explore in future work are how the thick CBD and the surrounding torus interact with the central binary, and whether the resulting orbital decay proceeds mainly through steady disk--binary interaction or requires transient events driven by disk instabilities.

One possible consequence of the formation of a CBD is episodic orbital decay. FU~Orionis outbursts have been proposed to be triggered by interactions between a binary and its surrounding disk~\citep[e.g.,][]{1992ApJ...401L..31B}. In a CEE CBD, an analogous episode, in which material suddenly falls into the binary cavity, possibly driven by the same disk instabilities often associated with FU~Ori events, would increase the drag on the central binary, producing a burst of orbital decay together with a transient brightening. Given the large masses of the CBDs formed in our simulations, we expect that, in some regions of parameter space, these disks may become gravitationally unstable, which we will investigate in future work. Such episodes could explain the multiple discrete ejection events seen in some PNe~\citep[such as the Butterfly Nebula,][]{2023ApJ...957...54B} and possibly the successive re-brightenings observed in some LRNe, such as V4332~Sgr~\citep{1999AJ....118.1034M,2026ApJ...999L..35B}. Interestingly, similarities between the emission of FU~Ori events and LRNe have previously been noted~\citep{2026arXiv260517005K}. As the system evolves, the disk is expected to transition from thick to thin on timescales of a few centuries for massive post-CEE binaries~\citep{2024A&A...688A..87W}, although this may differ for the low-mass systems studied here. At that point, the evolution of the central binary may be better described by the classical CBD framework already discussed in the CEE context~\citep[e.g.,][]{2011MNRAS.417.1466K,2024A&A...688A..87W,2024A&A...688A.128V}.

\begin{acknowledgements}

The authors thank Noam Soker for the comments and feedback about the paper. SVB acknowledges support from the NASA ATP program (NNH23ZDA001N-ATP) and the College of Letters and Science at UWM through the Chancellor's Graduate Student Award and the Research Excellence Award. PC acknowledges support from the NASA ATP program through NASA grant NNH17ZDA001N-ATP and National Science Foundation (NSF) via AST-2108269 and AST-2307885.

\end{acknowledgements}

\appendix

\section*{Data Availability}

The following data from the runs Q025, Q050, Q075, Q100, Q050-NCR, and Q050-T100 are available at \url{https://doi.org/10.5281/zenodo.23444527}:

\begin{itemize}
\item  \texttt{.dat} files containing the radial evolution of the height, midplane density, and midplane temperature;
\item  \texttt{.mov} videos showing the density and temperature evolution.
\end{itemize}

\texttt{MANGA} is available at:

\url{https://github.com/N-BodyShop/changa}.

The initial conditions to reproduce the numerical simulations as well as extra information about the results are available upon reasonable request. To evolve the star with MESA, we used version r23.05.1 and the example "1M\_pre\_ms\_to\_wd" from the test suite (adjusting the ZAMS mass for the 2M$_{\odot}$ RGB case).

\section{The Impact of numerics on the formation of a CBD} \label{sec:solver}
\subsection{Riemann Solver}

\begin{figure*}[htbp]
\parbox{\textwidth}{%
\includegraphics[width=0.95\textwidth]{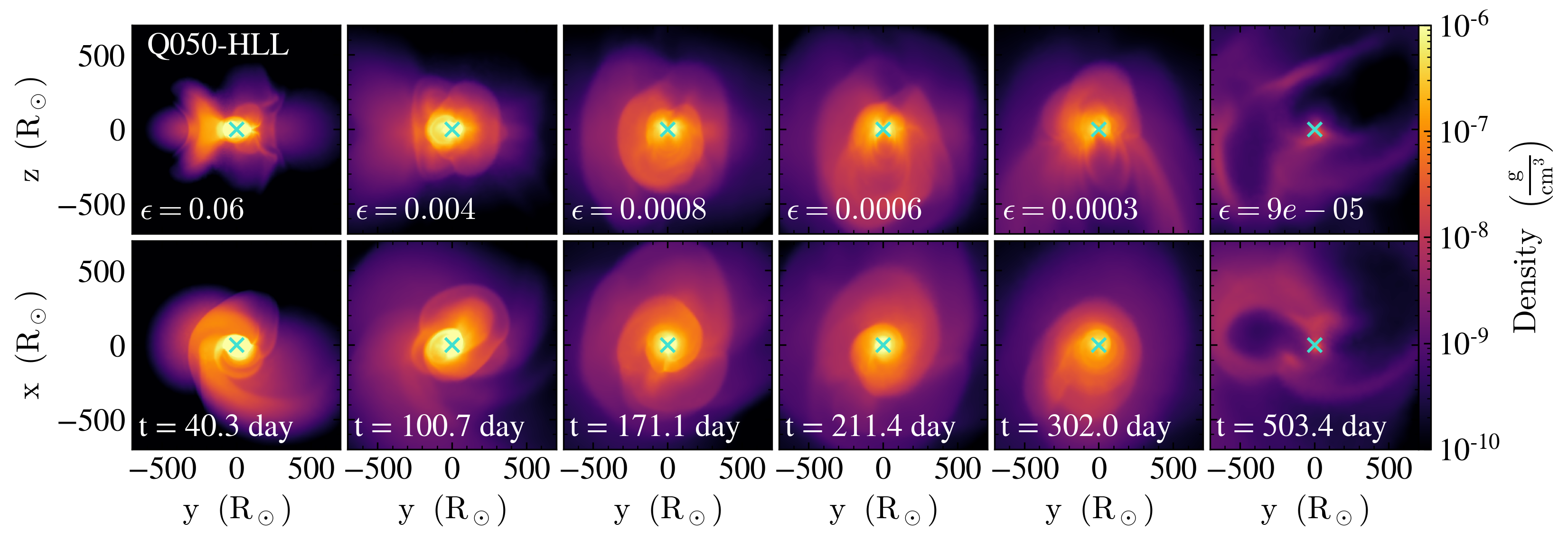}\\
}

\caption{Density distribution for a $2~\text{M}_\odot$ RGB star in a binary system with mass ratio $q=0.5$ for the Q050-HLL run, with the \texttt{HLL} Riemann solver. Each panel shows both $y$-$x$ and $y$-$z$ views at times $t = 40.3$, $100.7$, $171.1$, $211.4$, $302$, and $503.4$ days. Each panel has the $\epsilon$ at that given time. The cyan '$\times$' marks the center of mass of the core+companion duo.} 
\label{fig:projection_plots_HLL}
\end{figure*}

 \begin{figure*}[htbp]
\includegraphics[width=0.45\textwidth]{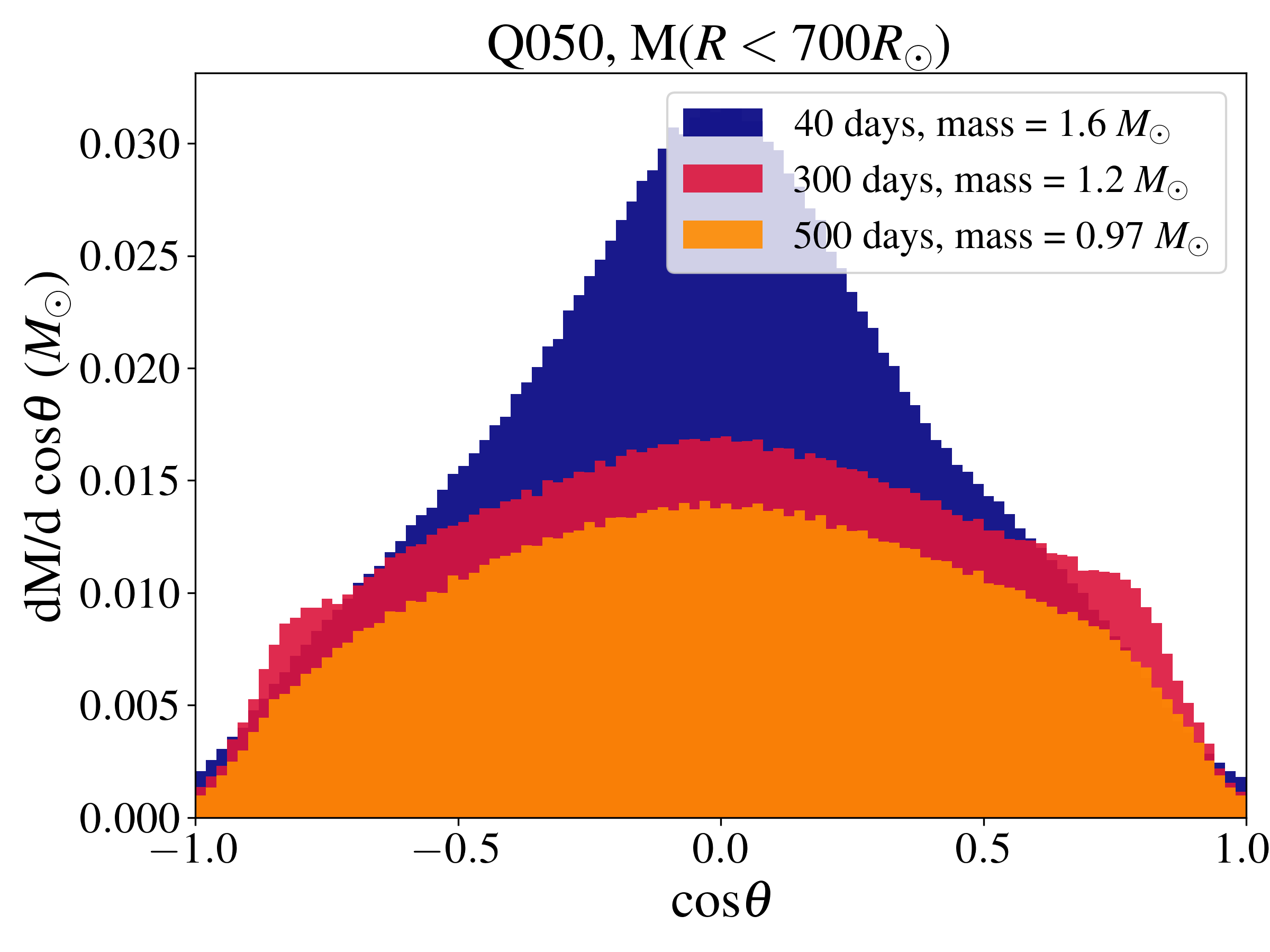}
\includegraphics[width=0.45\textwidth]{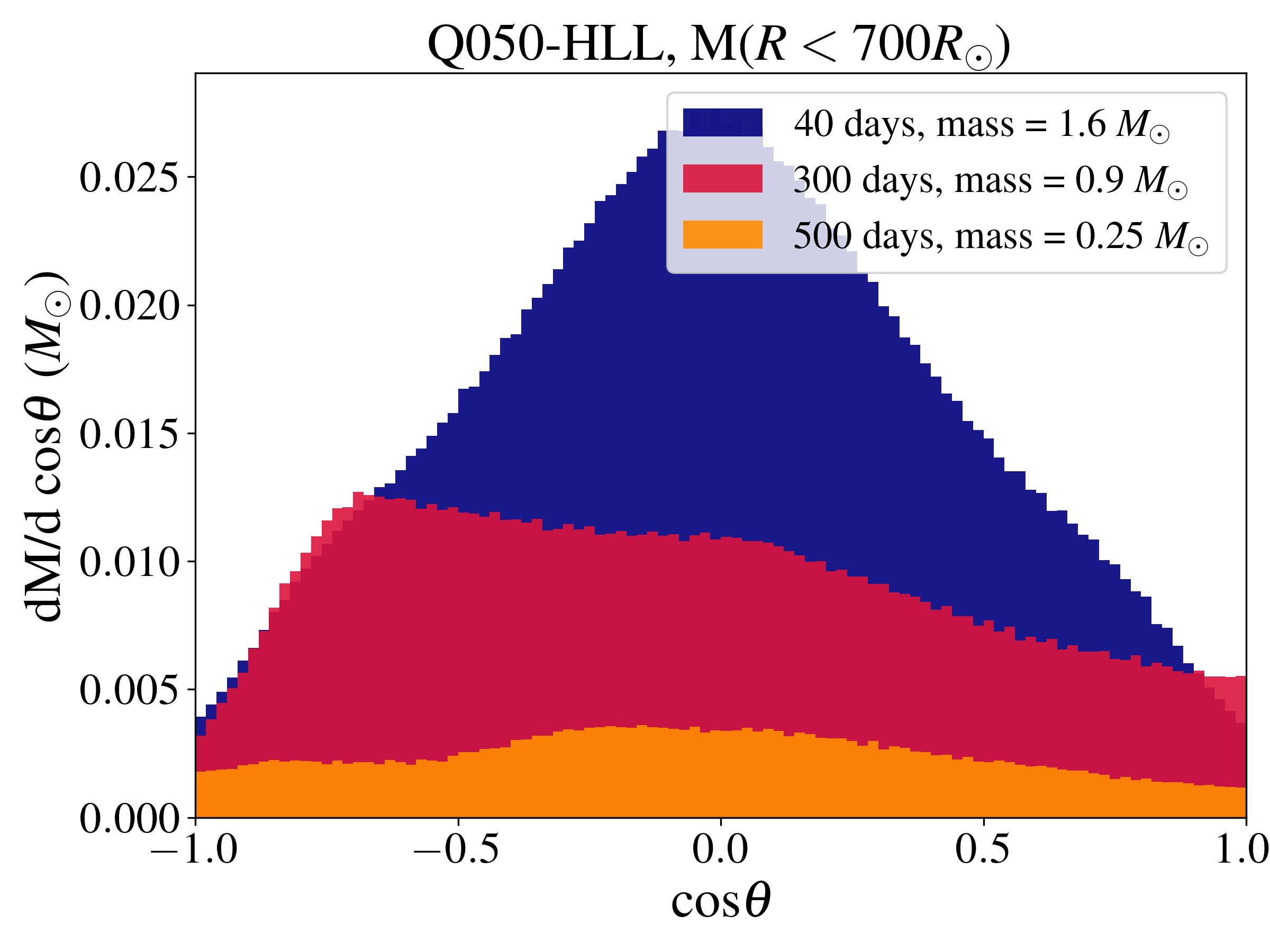}
\caption{Envelope mass distribution per $\cos(\theta)$ for a radius of 700~$R_{\odot}$ around the center of mass of core+companion duo for the 2 M$_\odot$ RGB star, q = 0.5, at 40 (blue), 300 (red) and 500 (orange) days. Left panel: Q050 run, \texttt{HLLC} solver. Right panel: Q050-HLL run, \texttt{HLL} solver.} 
\label{fig:histogram}

\end{figure*}

\begin{figure}
\includegraphics[width=0.47\textwidth]{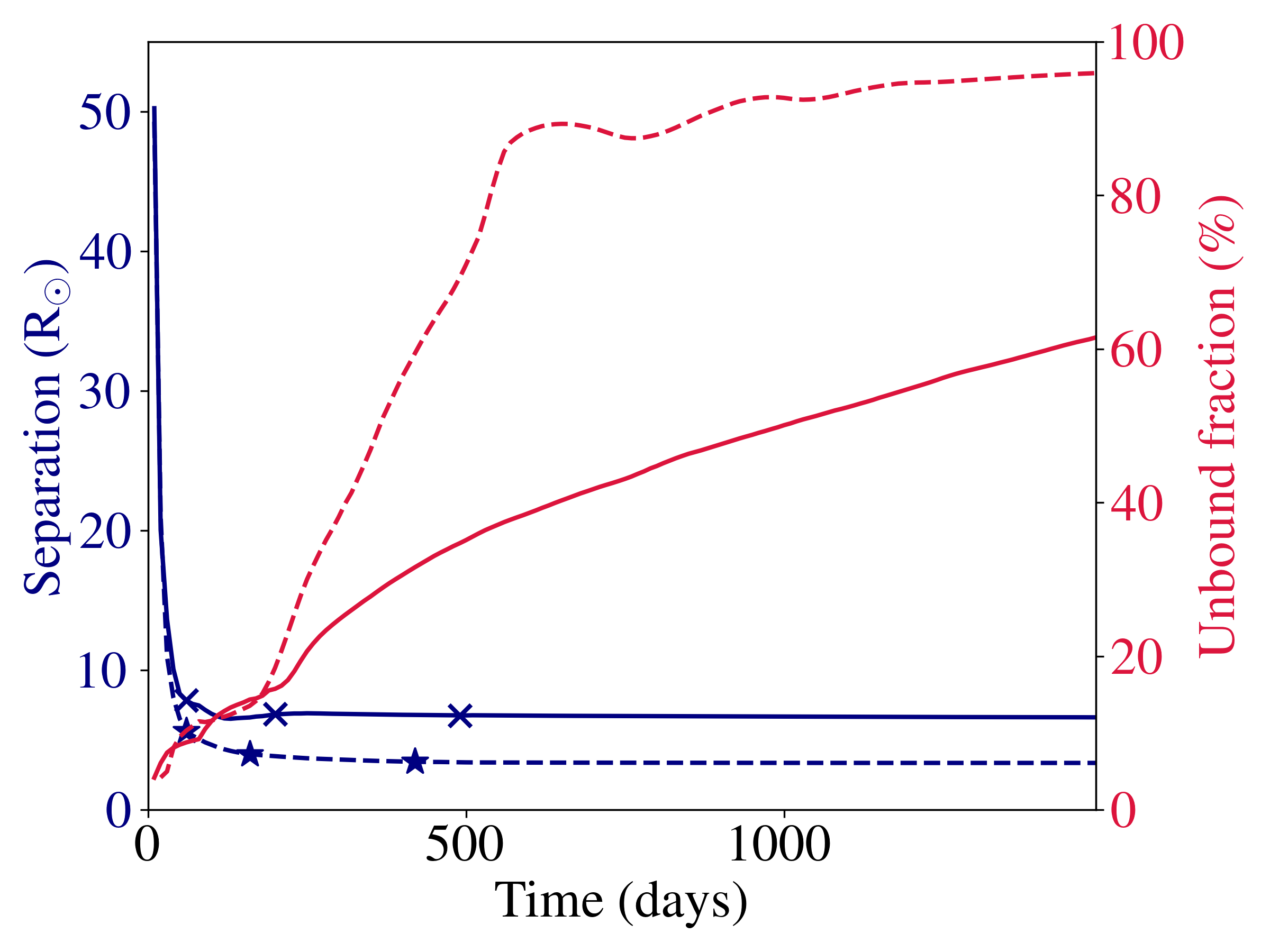}
\caption{Evolution of the unbound fraction (red) and orbital separation (blue) for a $2~M_{\odot}$ RGB star with $q=0.5$. Solid and dashed lines represent simulations using the \texttt{HLLC} (Q050 run) and \texttt{HLL} (Q050-HLL run) Riemann solvers, respectively. The markers 'x' and '*' indicate the times at which $\epsilon=0.01$, $0.001$, and $0.0001$ are reached for the \texttt{HLLC} and \texttt{HLL} runs, respectively. This reflects why we believe that $\epsilon=0.01$ is not a good threshold for the final separation at the plunge, given that at this moment the core+companion binary has not yet reached the plateau in the final separation right after the plunge.}
\label{fig:unbound_sep}
\end{figure}

Whether a CEE simulation forms a long-lived CBD and collimated outflows has proven sensitive to numerics. A decade ago, those structures were rarely reported in simulations of CEE. More recently, simulations increasingly started to find them \citep[e.g.,][]{2018MNRAS.480.1898C, 2019MNRAS.484..631R, 2024A&A...691A.244V, 2025A&A...697A..68G}. Over this period, resolution, solvers, and the treatment of softening length have all improved.  \citet{2025A&A...697A..68G} isolated one of these factors, showing that the softening length controls whether a cavity clears around the core and companion and whether collimated outflows form. Here we discuss another: the choice of Riemann solver, especially for our code \texttt{MANGA}.

This test is also motivated by \citet{2023MNRAS.526.5365V}, who simulated a similar system with the same code (\texttt{MANGA}) but with the \texttt{HLL} solver, and found no bipolar collimated outflows. In Figs.~\ref{fig:projection_plots} and~\ref{fig:projection_plots_HLL}, we show the projected density for runs Q050 (\texttt{HLLC}) and Q050-HLL (\texttt{HLL}) at $t = 40.3$, $100.7$, $171.1$, $211.4$, $302$, and $503.4$ days. In the edge-on ($y$–$z$) view, the \texttt{HLLC} run develops a bipolar structure, whereas the \texttt{HLL} run does not. In the \texttt{HLL} run, one side briefly develops a bipolar outflow, but this structure is disrupted and does not persist.

Fig.~\ref{fig:histogram} shows the envelope mass distribution as a function of $\cos\theta = z/r$ within $700\,R_\odot$ at three different times. The \texttt{HLLC} run remains approximately axisymmetric throughout the evolution, whereas the \texttt{HLL} run becomes asymmetric. In both runs the mass near the binary decreases with time, but much faster with \texttt{HLL}: by 500 days, $0.97\,M_\odot$ remains within $700\,R_\odot$ in the \texttt{HLLC} run, compared with only $0.25\,M_\odot$ in the \texttt{HLL} run.

This difference is connected to the energy transferred to the envelope. Fig.~\ref{fig:unbound_sep} shows the evolution of the orbital separation and of the unbound fraction for both solvers. The unbound fraction is computed from the sum of the mechanical ($E_{\rm mech}$) and internal ($E_{\rm int}$) energies of each particle $i$,
\begin{equation}
   E_{\rm tot} = E_{\rm mech} + E_{\rm int}.
\end{equation}
Particles with $E_{{\rm tot},i} < 0$ are considered bound, and those with $E_{{\rm tot},i} > 0$ unbound \citep[for details, see][]{2026ApJ...998...34B, 2019MNRAS.486.5809P}. As discussed in \citet{2026ApJ...998...34B}, the two solvers produce significantly different final separations: $3.8\,R_\odot$ with \texttt{HLL} versus $6.9\,R_\odot$ with \texttt{HLLC}. The tighter \texttt{HLL} binary deposits more orbital energy into the envelope and unbinds it faster: by the end of the simulation, nearly $100\%$ of the envelope is unbound with \texttt{HLL}, compared with slightly above $60\%$ with \texttt{HLLC}. In the \texttt{HLL} run, the envelope is removed too quickly for long-lived collimated structures to develop. In the \texttt{HLLC} run, a substantial amount of mass remains in the equatorial region, where it can reorganize into a CBD that collimates and feeds the polar outflows.

We attribute these differences to how each solver treats contact discontinuities. \texttt{HLLC} resolves them, whereas the more diffusive \texttt{HLL} smears them out. This matters at two stages. During the plunge, \texttt{HLLC} better resolves the wakes around the point masses, giving a more accurate drag and therefore a larger final separation. Later, it better maintains the density contrast between the disk and the polar regions at the base of the outflows, which allows the outflows to persist. Although the choice of Riemann solver applies only to grid-based codes (including AMR and moving-mesh codes), the underlying issue is how well density gradients and shocks are resolved. We therefore expect SPH simulations with insufficient resolution or too dissipative artificial viscosity to face analogous difficulties, although we have not tested this directly.

 A few reasons lead us to adopt the \texttt{HLLC} runs as our fiducial models. First, \texttt{HLLC} captures contact discontinuities that \texttt{HLL} does not, and these affect both the drag during the plunge and the structure of the outflows. Second, the CBD found with \texttt{HLLC} is consistent with recent simulations performed with independent codes and numerical methods \citep[e.g.,][]{2018MNRAS.480.1898C, 2019MNRAS.484..631R, 2024A&A...691A.244V, 2025A&A...697A..68G}, and the historical trend toward finding CBDs as numerics have improved over the years. Finally, the resulting multi-component morphology and collimated outflows are consistent with those observed in WFs and post-CEE PNe (that are confirmed or proposed CEE observations independently of the bipolar morphology, Sec.~\ref{sec:observations}). More broadly, these results show that both the final separation and the ejecta morphology of CEE simulations depend sensitively on numerics, and that numerical choices should be made with care.

\subsection{Resolution}
\label{sec:resolution}

Table~\ref{tab:our_simulations} shows simulations~(Q050-0.6M, Q050, Q050-1.6M, Q050-2.1M) with different resolutions~(0.6, 1.2, 1.6, and 2.1 million star particles) for otherwise identical initial conditions. Although some differences are observed in the final separation, the overall behavior of the inspiral remains unchanged. In particular, the final separation is very similar between the fiducial and 1.6M runs (6.9 and 7.1~R$_\odot$, respectively), while the 0.6M run yields a separation that is approximately 20\% larger, and the 2.1M run one that is $\sim13$\% smaller. The unbound fraction at 500~days is also similar between the fiducial and higher-resolution runs: 35\% for the fiducial run and 32\% for both the 1.6M and 2.1M runs, whereas the 0.6M run reaches 45\%. The outer edge of the Rayleigh-stable disk at 500~days is $\sim 225$, 237, and 298~R$_\odot$ for the fiducial, 1.6M, and 2.1M runs, respectively, comparable to the spread between Q050 and Q050-T100 (Sec.~\ref{sec:disk_size}).

\begin{figure*}
    \centering
    \includegraphics[width=0.85\linewidth]{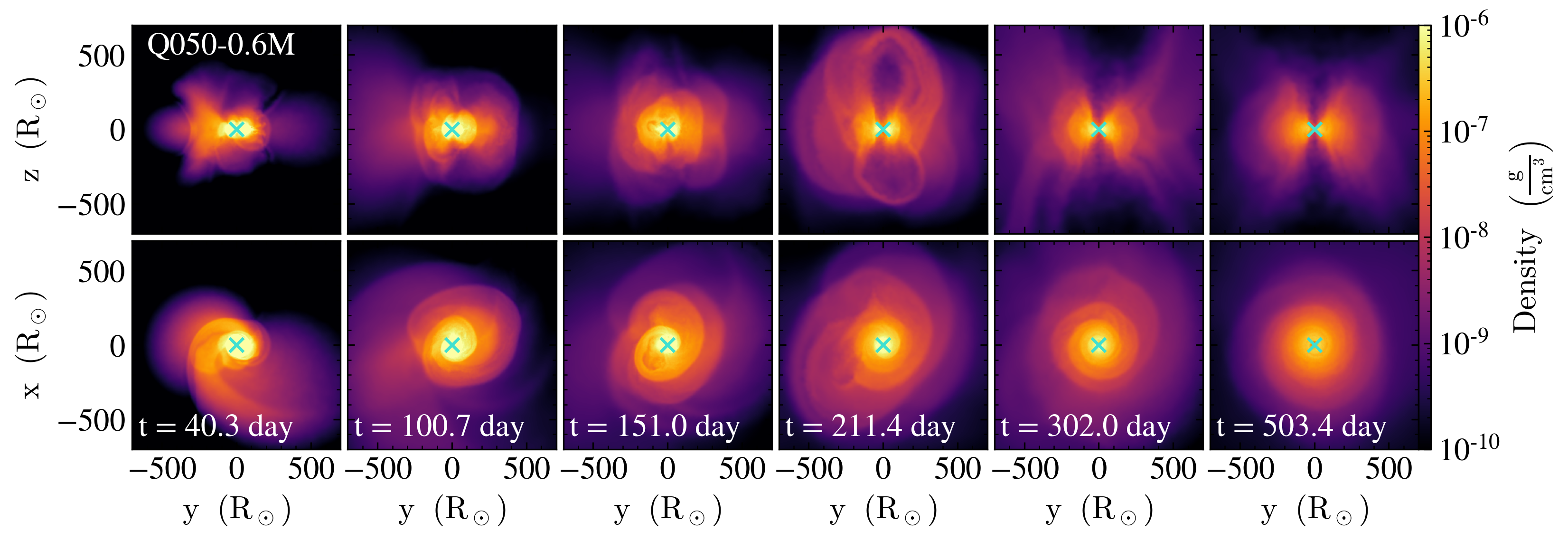}
    \includegraphics[width=0.85\linewidth]{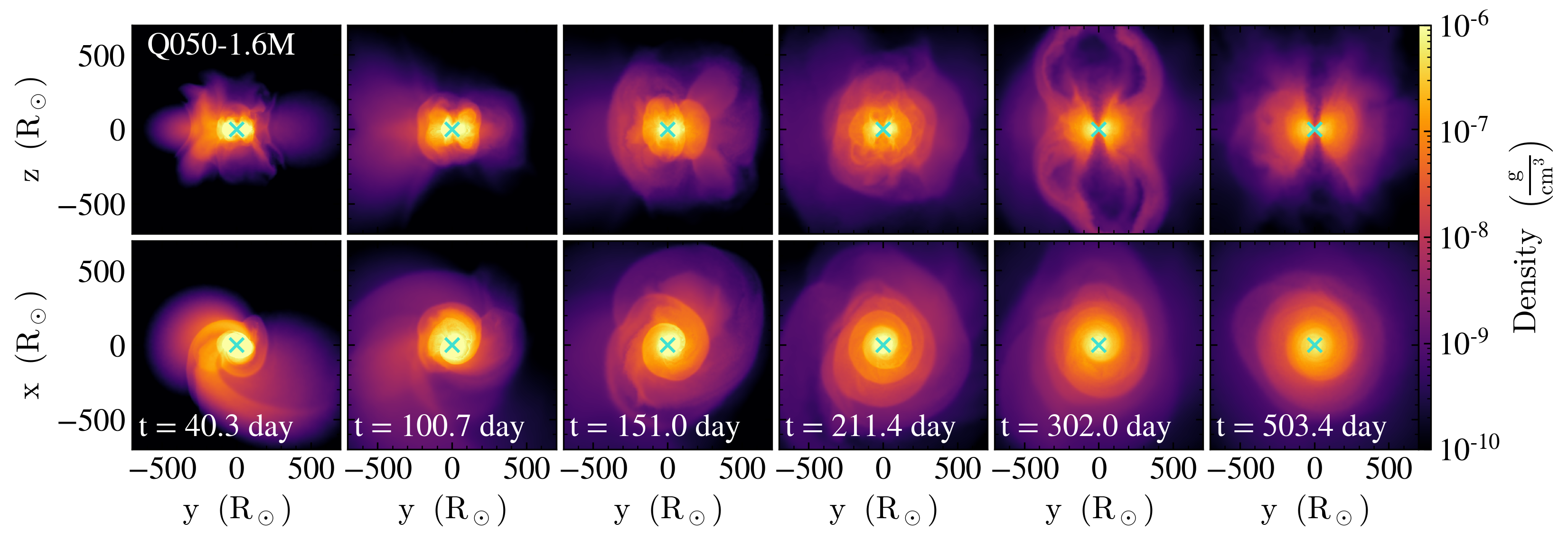}
        \includegraphics[width=0.85\linewidth]{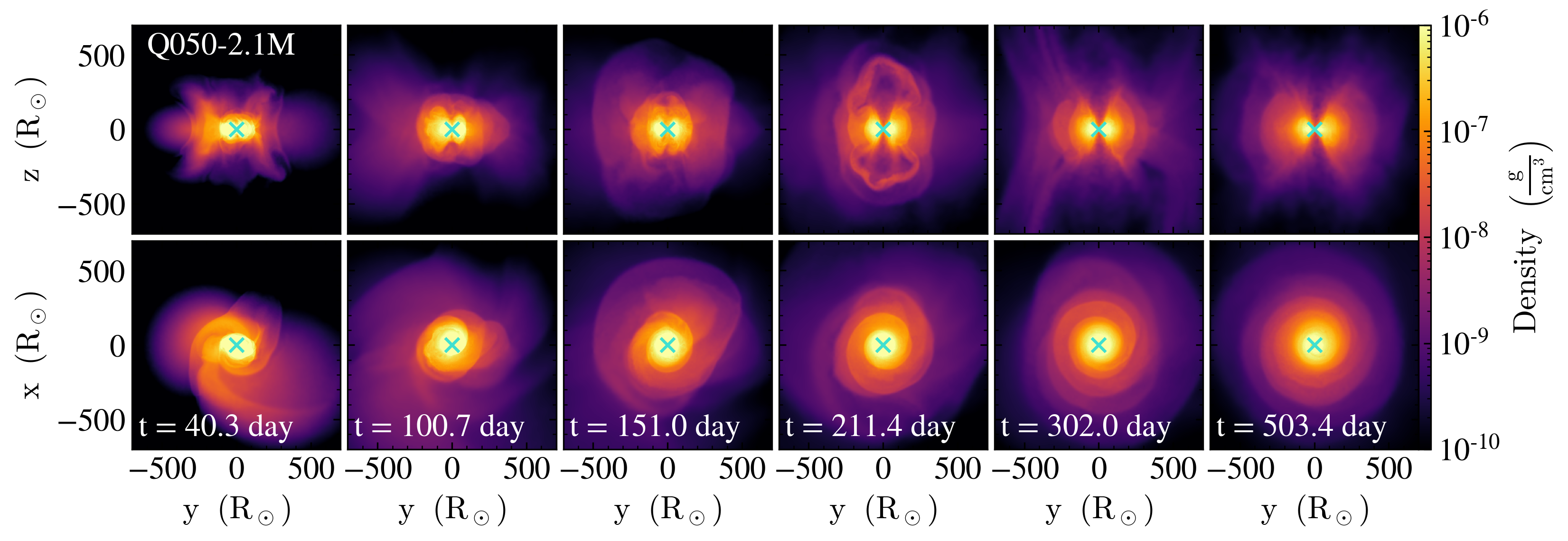}
\caption[Density distribution for a $2~\text{M}_\odot$ RGB star in a binary system with mass ratio $q=0.5$.]{
Density distribution for a $2~\text{M}_\odot$ RGB star in a binary system with mass ratio $q=0.5$. 
Top: low-resolution run (Q050-0.6M); middle: higher-resolution run (Q050-1.6M); bottom: highest-resolution run (Q050-2.1M). Each panel shows both $y$–$x$ and $y$–$z$ projections at times $t = 40.3$, $100.7$, $151$, $211.4$, $302$, and $503.4$ days. 
The cyan ‘$\times$’ marks the center of mass of the core–companion system. The low-resolution (Q050-0.6M), fiducial (Q050, 1.2 million particles; see Fig. 1), and highest-resolution (Q050-2.1M) runs develop collimated outflows at approximately the same time ($\sim200 $~days), whereas in the higher-resolution run (Q050-1.6M) they appear roughly 100 days later.
}
    \label{fig:resolution}
\end{figure*}

In terms of the morphology of the ejecta, the resolution primarily affects the timing at which collimated outflows appear, but not their overall presence (see Fig.~\ref{fig:resolution}, for the lower and higher resolution cases, and Fig.~\ref{fig:projection_plots}, for the fiducial case). In all cases, a substantial equatorial mass, consistent with a CBD and torus, is present, along with collimated outflows bursting from the central region. This indicates that while some features of the flow, such as the onset of outflows, are resolution-dependent, the global morphology seems to be robust under the tested resolutions (that varies from 0.6 to 2.1 million star particles). Moreover, neither the onset time of the outflows nor the final separation changes monotonically with resolution.

\bibliography{apssamp}

\end{document}